\documentclass[journal]{./sty/IEEEtran}

\usepackage[pdftex]{graphicx,color}
\usepackage{latexsym,amssymb,xfrac,ifthen,cite,array,url}
\usepackage[cmex10]{amsmath}
\usepackage{array}
\usepackage{fixltx2e}
\graphicspath{{pdf/}}

\newcommand{\ignore}[1]{}

\newcommand{\Fig}[1]{Figure~\ref{#1}}

\newcommand{\eqdef}{\stackrel{\scriptscriptstyle\bigtriangleup}{=} }
\newcommand{\vct}[1]{\mathbf{#1}}    

\newcommand{\restrict}[2]{\left.#1\right|_{#2}}

\newcommand{\diag}{\operatorname{diag}}

\newcommand{\cent}[1]{\makebox(0,0){#1}}
\newcommand{\pos}[2]{\makebox(0,0)[#1]{#2}}

\newcommand{\signame}[1]{#1}

\newcommand{\T}{\mathsf{T}}
\renewcommand{\H}{\mathsf{H}}

\newcommand{\calB}{\mathcal{B}}

\newcommand{\calN}{\mathcal{N}}

\newcommand{\E}{\operatorname{E}}

\makeatletter
\DeclareFontFamily{U}{MnSymbolA}{}
\DeclareSymbolFont{MnSyA}{U}{MnSymbolA}{m}{n}
\DeclareFontShape{U}{MnSymbolA}{m}{n}{
<-6> MnSymbolA5
<6-7> MnSymbolA6
<7-8> MnSymbolA7
<8-9> MnSymbolA8
<9-10> MnSymbolA9
<10-12> MnSymbolA10
<12-> MnSymbolA12}{}
\DeclareMathSymbol{\smallrightarrow}{\mathrel}{MnSyA}{0}
\DeclareMathSymbol{\smallleftarrow}{\mathrel}{MnSyA}{2}
\DeclareMathSymbol{\smallleftrightarrow}{\mathrel}{MnSyA}{16}
\newcommand{\smallrightarrowfill@}{\arrowfill@\relbar\relbar\smallrightarrow}
\newcommand{\smallleftarrowfill@}{\arrowfill@\smallleftarrow\relbar\relbar}
\newcommand{\smallleftrightarrowfill@}
{\arrowfill@\smallleftarrow\relbar\smallrightarrow}
\renewcommand{\overrightarrow}{\mathpalette{\overarrow@\smallrightarrowfill@}}
\renewcommand{\overleftarrow}{\mathpalette{\overarrow@\smallleftarrowfill@}}
\renewcommand{\overleftrightarrow}
{\mathpalette{\overarrow@\smallleftrightarrowfill@}}
\makeatother

\providecommand{\msgf}[2]{\protect\overrightarrow{#1}_{\mspace{-3mu}#2}} 
\providecommand{\msgb}[2]{\protect\overleftarrow{#1}_{\mspace{-3mu}#2}} 

\providecommand{\vmsgf}[2]{\protect\overrightarrow{\vct{#1}}_{#2}}
\providecommand{\smsgf}[2]{\protect\overrightarrow{\tilde{{#1}}}_{#2}}
\providecommand{\vmsgb}[2]{\protect\overleftarrow{\vct{#1}}_{#2}}
\providecommand{\smsgb}[2]{\protect\overleftarrow{\tilde{{#1}}}_{#2}}

\definecolor{gray}{rgb}{0.5, 0.5, 0.5}

\definecolor{lightgray}{rgb}{0.7, 0.7, 0.7}

\newcounter{examplecntr}
\renewcommand{\theexamplecntr}{\arabic{examplecntr}}
{\begin{trivlist}\small\item[]\refstepcounter{examplecntr}%
 {\bfseries Example~\theexamplecntr%
  \ifthenelse{\equal{#1}{}}{ }{ (#1) } 
}}%
{\hfill$\Box$\end{trivlist}}

\newcommand{\knownBox}{\cent{\rule{1.75\unitlength}{1.75\unitlength}}}

\newcommand{\connectionDot}{\circle*{1}}

\newcommand{\plusSign}{%
\begin{picture}(0,0)(0,0)
\put(0,0){\circle{4}}
\put(-1,0){\line(1,0){2}}
\put(0,-1){\line(0,1){2}}
\end{picture}
}

\newcommand{\threshold}{%
   \begin{picture}(0,0)(0,0)
     \thicklines
     \put(0,-1.5){\line(-1,0){2.2}}
     \put(0,-1.5){\line(0,1){3}}
     \put(0,1.5){\line(1,0){2.2}}
   \end{picture}
}

\newcommand{\flashADC}{%
   \begin{picture}(0,0)(0,0)
     \thicklines
     \put(-2,-3){\line(-1,0){2}}
     \put(-2,-1){\line(0,-1){2}}
     \put(0,-1){\line(-1,0){2}}
    \put(0,-1){\line(0,1){2}}    
     \put(0,1){\line(1,0){2}}
     \put(2,1){\line(0,1){2}}
     \put(2,3){\line(1,0){2}}
   \end{picture}
}

\newcommand{\integratorBox}[2]{%
\begin{picture}(55,25)(0,-20)   
\put(0,0){\vector(1,0){13}}
\put(15,0){\plusSign}
\put(17,0){\vector(1,0){6.5}}
\put(27.5,0){\circle{8}}  \put(27.5,0){\cent{#1}}   
\put(31.5,0){\vector(1,0){6}}
\put(37.5,-5){\framebox(10,10){\Large$\int$}}
\put(47.5,0){\line(1,0){7.5}}
\put(15,-6.8){\vector(0,1){4.8}}
\put(15,-11.5){\circle{9}}  \put(15,-11.5){\cent{#2}}   
\put(15,-20){\vector(0,1){3.9}}
\put(32.5,-20){\line(-1,0){17.5}}
\put(32.5,-25){\framebox(10,10){}}
 \put(37.5,-20){\threshold}
\put(55,-20){\vector(-1,0){12.5}}
\put(55,0){\line(0,-1){7.5}}
\put(55,-12.5){\connectionDot}
 \put(55,-12.5){\line(1,2){2.5}}
\put(55,-12.5){\line(0,-1){7.5}}
\end{picture}
} 

\begin{document}

\title{Control-Bounded Analog-to-Digital Conversion:\\Transfer Function Analysis, Proof of Concept,\\
and Digital Filter Implementation}

\author{\IEEEauthorblockN{Hans-Andrea Loeliger, Hampus Malmberg, and Georg Wilckens}
\thanks{%
H.~A.\ Loeliger and H.~Malmberg are with
the Dept.\ of Information Technology and Electrical Engineering,
ETH Zurich, CH-8092 Z\"urich, Switzerland, Email: \{loeliger,malmberg\}@isi.ee.ethz.ch.
G.~Wilckens was also with the same Dept.\ of ETH Zurich;
he is now with Swiss International Airlines, Zurich, Email: georg.wilckens@gmail.com.
}}

\maketitle

\begin{abstract}
Control-bounded analog-to-digital conversion
has many commonalities with delta-sigma conversion,
but it can profitably use more general analog filters.
The paper describes the operating principle, 
gives a transfer function analysis,
presents a proof-of-concept implementation, 
and describes the digital filtering in detail.
\end{abstract}


\section{Introduction}
\label{sec:Intro}

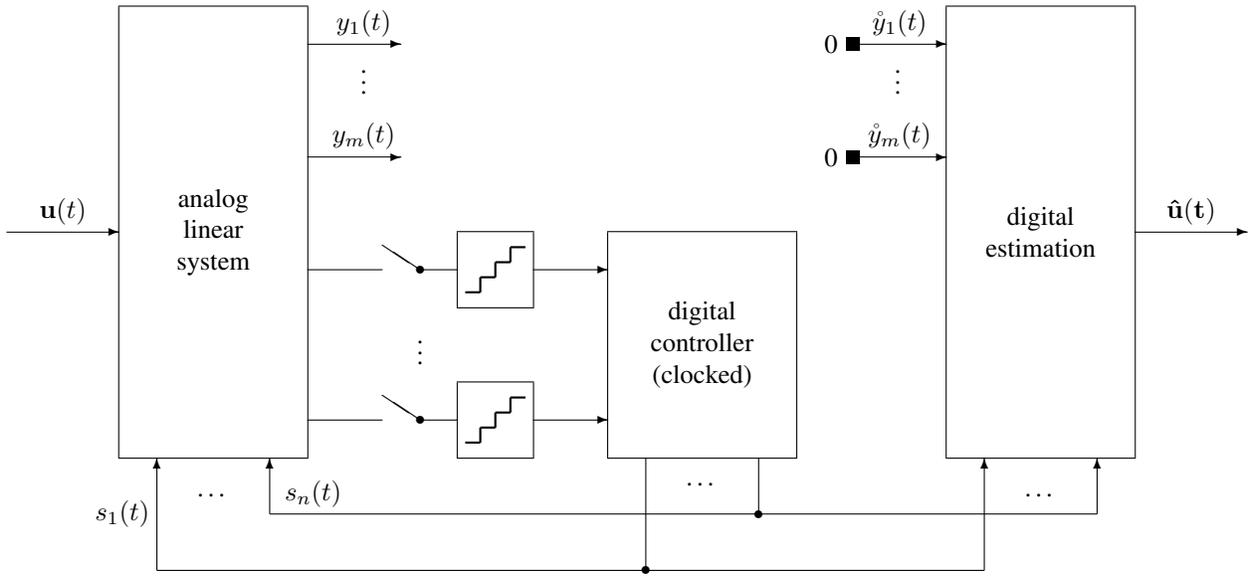
\begin{figure*}
\centering
\begin{picture}(165,75)(0,0)

\put(0,45){\vector(1,0){15}}  \put(7.5,46){\pos{cb}{$\vct u(t)$}}
\put(15,15){\framebox(25,60){\centering\parbox{25\unitlength}{\centering analog\\ linear\\ system}}}
\put(40,70){\vector(1,0){12.5}}  \put(47.5,71){\pos{cb}{$y_1(t)$}}
\put(47.5,66){\cent{$\vdots$}}
\put(40,55){\vector(1,0){12.5}}  \put(47.5,56){\pos{cb}{$y_m(t)$}}

\put(40,40){\line(1,0){10}}
\put(55,40){\line(-3,2){5}}
\put(55,40){\connectionDot}
\put(55,40){\line(1,0){5}}
\put(60,35){\framebox(10,10){}}
\put(65,40){\flashADC}
\put(70,40){\vector(1,0){10}}
\put(55,30){\cent{$\vdots$}}
\put(40,20){\line(1,0){10}}
\put(55,20){\line(-3,2){5}}
\put(55,20){\connectionDot}
\put(55,20){\line(1,0){5}}
\put(60,15){\framebox(10,10){}}
\put(65,20){\flashADC}
\put(70,20){\vector(1,0){10}}

\put(80,15){\framebox(25,30){\centering\parbox{25\unitlength}{\centering digital\\ controller\\ (clocked)}}}
\put(85,15){\line(0,-1){15}}
 \put(85,0){\connectionDot}
\put(85,0){\line(-1,0){65}}
\put(20,0){\vector(0,1){15}}  \put(19,7.5){\pos{cr}{$s_1(t)$}}
\put(85,0){\line(1,0){45}}
\put(130,0){\vector(0,1){15}}
\put(27.5,10){\cent{\ldots}}
\put(92.5,11.5){\cent{\ldots}}
\put(137.5,10){\cent{\ldots}}
\put(100,15){\line(0,-1){7.5}}
 \put(100,7.5){\connectionDot}
\put(100,7.5){\line(-1,0){65}}
\put(35,7.5){\vector(0,1){7.5}}  \put(37,8.5){\pos{bl}{$s_n(t)$}}
\put(100,7.5){\line(1,0){45}}
\put(145,7.5){\vector(0,1){7.5}}

\put(112.5,70){\vector(1,0){12.5}}  \put(118.75,71){\pos{cb}{$\mathring{y}_1(t)$}}
 \put(112.5,70){\knownBox}  \put(110.5,70){\pos{cr}{0}}
\put(118.5,66){\cent{$\vdots$}}
\put(112.5,55){\vector(1,0){12.5}}  \put(118.75,56){\pos{cb}{$\mathring{y}_m(t)$}}
 \put(112.5,55){\knownBox}  \put(110.5,55){\pos{cr}{0}}
\put(125,15){\framebox(25,60){\centering\parbox{25\unitlength}{\centering digital\\ estimation}}}
\put(150,45){\vector(1,0){15}}  \put(157.5,46){\pos{cb}{$\vct{\hat u(t)}$}}

\end{picture}
\caption{\label{fig:GeneralApproach}%
Control-bounded analog-to-digital converter.
}
\end{figure*}

Based on control-aided analog-to-digital conversion as in \cite{LBWB:ITA2011c}, 
control-bounded analog-to-digital conversion was proposed in \cite{LgW:ITA2015c}.
The general structure of such an analog-to-digital converter (ADC)
is shown in \Fig{fig:GeneralApproach}: 
the continuous-time analog input signal $u(t)$ 
(or $\vct u(t)$ as in (\ref{eqn:vctu})) 
is fed into an analog linear system/filter, which is subject 
to digital control. 
The digital control ensures that all analog quantities,
including, in particular, the signals $y_1(t)$, \ldots, $y_m(t)$, 
remain within their proper physical limits.
Using the digital control signals $s_1(t)$, \ldots, $s_n(t)$,
the digital estimation unit tracks the state of the analog system
and produces (arbitrarily spaced samples of) an estimate $\hat u(t)$ of $u(t)$.
The analog signals $y_1(t)$, \ldots, $y_m(t)$, which are not available to the digital estimator,
play a key role in the estimation as will be detailed in Section~\ref{sec:Theory}.

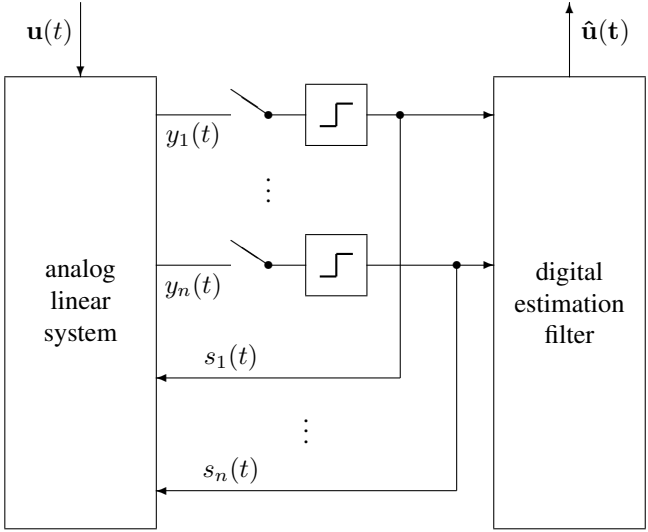
\begin{figure}
\centering
\begin{picture}(85,70)(0,0)

\put(10,70){\vector(0,-1){10}}  \put(9,66){\pos{cr}{$\vct u(t)$}}
\put(0,0){\framebox(20,60){\parbox{20\unitlength}{\centering analog\\ linear\\ system}}}

\put(20,55){\line(1,0){10}}    \put(25,54){\pos{ct}{$y_1(t)$}}
\put(35,55){\line(-3,2){5}}
\put(35,55){\connectionDot}
\put(35,55){\line(1,0){5}}
\put(40,51){\framebox(8,8){}}
 \put(44,55){\threshold}
\put(48,55){\vector(1,0){17}}
\put(52.5,55){\connectionDot}
\put(52.5,55){\line(0,-1){35}}

\put(35,46){\cent{$\vdots$}}

\put(20,35){\line(1,0){10}}    \put(25,34){\pos{ct}{$y_n(t)$}}
\put(35,35){\line(-3,2){5}}
\put(35,35){\connectionDot}
\put(35,35){\line(1,0){5}}
\put(40,31){\framebox(8,8){}}
 \put(44,35){\threshold}
\put(48,35){\vector(1,0){17}}
\put(60,35){\connectionDot}
\put(60,35){\line(0,-1){30}}

\put(52.5,20){\vector(-1,0){32.5}}   \put(30,21){\pos{cb}{$s_1(t)$}}
\put(40,14){\cent{$\vdots$}}
\put(60,5){\vector(-1,0){40}}    \put(30,6){\pos{cb}{$s_n(t)$}}

\put(75,60){\vector(0,1){10}}  \put(76.5,66){\pos{cl}{$\vct{\hat u(t)}$}}
\put(65,0){\framebox(20,60){\parbox{20\unitlength}{\centering digital\\ estimation\\ filter}}}

\end{picture}
\caption{\label{fig:GenSystemSimpleControl}%
A special case of \Fig{fig:GeneralApproach} 
where the control is effected by binary feedback
from the signals $y_1,\ \ldots,\ y_n$, 
which include the control bounded signals $y_1,\ \ldots,\ y_m$.
}
\end{figure}

In the important special case shown in \Fig{fig:GenSystemSimpleControl}, 
the $\{ +1, -1\}$-valued control signals $s_1(t)$, \ldots, $s_n(t)$
are obtained by sampling and thresholding 
the analog signals $y_1,\ \ldots,\ y_n$,
which include the control-bounded signals $y_1(t)$, \ldots, $y_m(t)$, $m\leq n$.

As may be conjectured from \Fig{fig:GenSystemSimpleControl}, 
control-bounded converters may be viewed as generalizations
of delta-sigma ($\Delta\Sigma$) converters \cite{PST:2017}. 
Indeed, for \mbox{$n=1$}, a control-bounded converter as in \Fig{fig:GenSystemSimpleControl}
has no advantage over a standard $\Delta\Sigma$ converter.
For \mbox{$n \geq 2$}, however, control-bounded converters can 
use analog systems/filters that 
cannot be handled by conventional $\Delta\Sigma$ techniques.

The descriptions in \cite{LBWB:ITA2011c,LgW:ITA2015c}
are terse and may not be easily accessible to analog designers. 
Moreover, the transfer function analysis in \cite{LgW:ITA2015c} 
covers only the case $m=1$, 
the performance analysis in \cite{LgW:ITA2015c}
is rudimentary, and no measurements of a real circuit are reported.

In this paper, we describe the operating principle 
and the digital estimation filter in more detail,
we give a full transfer function analysis,
and we report measurements of a breadboard circuit prototype.
In particular, this paper provides sufficient information for analog designers
to experiment with control-bounded ADCs. 

Much space will be given to the analysis of a single example:
digital control, noise and mismatch properties, simulations,
and measurements of the hardware prototype. 
This example%
---a chain of integrators as in \cite{LgW:ITA2015c}---%
closely resembles
a multi-stage noise shaping (MASH) $\Delta\Sigma$
ADC \cite{ASS:sdadc1996,R:sdto2011}, 
but with an analog part that precludes
a conventional digital cancellation scheme.
(Other analog circuit topologies with attractive properties 
will be described elsewhere.)

The paper is structured as follows. 
The operating principle and the basic transfer function analysis 
of control-bounded converters are given in Section~\ref{sec:Theory}.
A~conversion noise analysis is given in Section~\ref{sec:PerformanceAnalysis}.
The circuit example is presented and analyzed in Section~\ref{sec:Example}.
Some enhancements (including, in particular, a tailored dithering method) 
are discussed in Section~\ref{sec:Improvements}.
The sensitivity to thermal noise and component mismatch is considered in 
Section~\ref{sec:ThermalNoiseMismatch}.
The digital estimation filter is described 
in Section~\ref{sec:Computation}. 
The actual derivation of this filter is outlined in the Appendix.

\section{Operating Principle} 
\label{sec:Theory}

\subsection{Analog Part and Digital Control}

Consider the system of \Fig{fig:GeneralApproach}. 
The continuous-time input signal $u(t)$ is assumed to be bounded, 
i.e.,  $|u(t)| \leq b_{\signame{u}}$ for all times $t$.
More generally, the input signal may be a vector
\begin{equation} \label{eqn:vctu}
\vct{u}(t) \eqdef \big( u_1(t),\ \ldots,\ u_k(t) \big)^\T
\end{equation}
with bounded components 
$|u_\ell(t)| \leq b_{\signame{u}}$ for all $t$ and $\ell=1,\ldots,k$.
The analog linear system produces 
a continuous-time vector signal
\begin{equation}
\vct{y}(t) \eqdef \big( y_1(t),\ \ldots,\ y_m(t) \big)^\T,
\end{equation}
and the digital control in \Fig{fig:GeneralApproach}
ensures that 
\begin{equation} \label{eqn:allybound}
|y_\ell(t)| \leq b_{\signame{y}} \text{~~~ for all $t$ and $\ell=1,\ldots,m$.}
\end{equation}
We also assume that the digital control is additive,
i.e., 
\begin{equation} \label{eqn:neteffectOfControl}
\vct{y}(t) = \breve{\vct{y}}(t) - \vct{q}(t),
\end{equation}
where $\breve{\vct{y}}(t)$ 
(given by (\ref{eqn:UncontrolledOutput}) below)
is the fictional signal $\vct{y}(t)$ 
that would result without the digital control
and where $\vct{q}(t)$ is fully determined by the control signals 
$s_1(t)$, \ldots, $s_n(t)$.

At this point, we have already finished the discussion of the digital control 
in this section: its role and its effect are fully described by 
(\ref{eqn:allybound}) and (\ref{eqn:neteffectOfControl}).

Clearly, neither $\breve{\vct{y}}(t)$ nor $\vct{q}(t)$ are bounded by $b_{\signame{y}}$.
In fact, the first key idea of control-bounded conversion 
is to use $\vct{q}(t)$ as a proxy for $\breve{\vct{y}}(t)$,
and this approximation will be good only if the magnitude of $\breve{\vct{y}}(t)$ 
is much larger than the magnitude of $\vct{y}(t)$ 
(as will be made precise below).
Note that $\vct{q}(t)$ may be very complicated,
but it is, in principle, known to the digital estimator
since $\vct{q}(t)$ is fully determined by $s_1(t)$, \ldots, $s_n(t)$.

We now assume that the uncontrolled analog filter is time-invariant and stable%
\footnote{The extension of the following transfer function analysis to unstable
analog systems is possible, but beyond the scope of this paper.} 
with impulse response matrix 
\begin{IEEEeqnarray}{rCl} \label{eq:impulseResponseMatrix}
\vct{g}(t) &\eqdef& \begin{pmatrix}g_{1,1}(t) & \dots & g_{1, k}(t) \\
  \vdots & \ddots & \vdots \\
  g_{m,1}(t) & \dots & g_{m,k}(t)
\end{pmatrix},
\end{IEEEeqnarray} where $g_{i,j}(t)$ is the impulse response 
from $u_j(t)$ to $y_i(t)$. 
We then have
\begin{IEEEeqnarray}{rCl} \label{eqn:UncontrolledOutput}
  \breve{\vct{y}}(t) & = & (\vct{g} \ast \vct{u})(t) \\
  & \eqdef & \begin{pmatrix}
    (g_{1,1} \ast u_1)(t) + \ldots + (g_{1,k} \ast u_k)(t) \\
    \vdots \\
    (g_{m,1} \ast u_1)(t) + \ldots + (g_{m,k} \ast u_k)(t)
  \end{pmatrix}.
\end{IEEEeqnarray}

We will also need the (elementwise) Fourier transform of (\ref{eq:impulseResponseMatrix}),
which will be denoted by $\vct{G}(\omega)$ and will be called analog transfer function (ATF) matrix.

\subsection{Digital Estimation and Transfer Functions}
\label{sec:EstimTFA}

Using the impulse response matrix $\vct{h}$ 
defined in (\ref{eqn:EstimFilterDef}) below,
we define the continuous-time estimate
\begin{equation} \label{eqn:DefHatu}
\hat{\vct{u}}(t) \eqdef (\vct{h} \ast \vct{q})(t),
\end{equation}
which can be written as
\begin{IEEEeqnarray}{rCl}
\hat{\vct{u}}(t) 
  & = & (\vct{h} \ast \breve{\vct{y}})(t) - (\vct{h} \ast \vct{y})(t) \label{eqn:EstimDecomp} \IEEEeqnarraynumspace\\
  & \approx &  (\vct{h} \ast \breve{\vct{y}})(t) \label{eq:KeyApprox}\\
  & = &  (\vct{h} \ast \vct{g} \ast \vct{u})(t). \label{eqn:SignalPath}
\end{IEEEeqnarray}
Note that the step from (\ref{eqn:EstimDecomp}) to (\ref{eq:KeyApprox})
uses the mentioned approximation $\breve{\vct{y}}\approx \vct{q}$,
or, equivalently, the approximation 
\begin{equation} \label{eqn:SetYZero}
\vct{y}(t) \approx \mathring{\vct{y}}(t) \eqdef \vct{0},
\end{equation}
as illustrated in \Fig{fig:GeneralApproach}.

The impulse response matrix $\vct{h}$  is determined
by its (elementwise) Fourier transform
\begin{IEEEeqnarray}{rCl} \label{eqn:EstimFilterDef}
\vct{H}(\omega) &\eqdef& {\vct{G}(\omega)}^\H\left(\vct{G}(\omega)\vct{G}(\omega)^\H + \eta^2\vct{I}_{m}\right)^{-1},
\end{IEEEeqnarray}
where $(\cdot)^\H$ denotes Hermitian transposition,
$\vct{I}_{m}$ is the $m$-by-$m$ identity matrix, 
and $\eta>0$ is a design parameter. 
Each element of $\vct{h}(t)$ is stable,
and arbitrarily spaced samples of (\ref{eqn:DefHatu})
can be computed from the control signals $s_1(t)$, \ldots, $s_n(t)$
as will be described in Section~\ref{sec:Computation}.

Equations (\ref{eqn:EstimDecomp}) and (\ref{eqn:SignalPath}) 
can then be interpreted as follows.
Eq.\ (\ref{eqn:SignalPath}) is the signal path:
the signal $\vct{u}(t)$ 
is filtered with the signal transfer function (STF) matrix
\begin{equation} \label{eqn:ADCTransferFunction}
\vct{H}(\omega)\vct{G}(\omega) = \vct{G}(\omega)^\H\left(\vct{G}(\omega)\vct{G}(\omega)^\H + \eta^2\vct{I}_{m}\right)^{-1}\vct{G}(\omega).
\end{equation}
The second term in (\ref{eqn:EstimDecomp}) is the 
conversion error
\begin{IEEEeqnarray}{rCl}
\epsilon(t) & 
\eqdef & \hat{\vct{u}}(t) - (\vct{h}\ast \vct{g}\ast \vct{u})(t) \label{eqn:CorrErrorSignal}\\
 & = & - (\vct{h} \ast \vct{y})(t) \label{eqn:ErrorSigY}
\end{IEEEeqnarray}
with $\vct{y}(t)$ bounded as in (\ref{eqn:allybound}).
Because of (\ref{eqn:ErrorSigY}), $\vct{H}(\omega)$ 
will be called noise transfer function (NTF) matrix.

In the important special case where $\vct{u}(t)$ is scalar (i.e., \mbox{$k=1$}), 
the ATF matrix $\vct{G}(\omega)$ is a column vector
and the NTF matrix (\ref{eqn:EstimFilterDef}) is a row vector.
Using the matrix inversion lemma, 
the latter
can be written as 
\begin{equation} \label{eqn:scalarinputNTF}
\vct{H}(\omega) = \frac{\vct{G}(\omega)^\H}{\|\vct{G}(\omega)\|^2 + \eta^2}
\end{equation}
and the STF matrix (\ref{eqn:ADCTransferFunction}) reduces to the scalar STF
\begin{equation} \label{eqn:scalarinputSTF}
\vct{H}(\omega) \vct{G}(\omega) = 
\frac{\|\vct{G}(\omega)\|^2}{\|\vct{G}(\omega)\|^2 + \eta^2}
\end{equation}
Note that (\ref{eqn:scalarinputSTF}) does not entail a phase shift
and is free of aliasing (hence the title of \cite{LgW:ITA2015c}):  
the sampling in \Fig{fig:GeneralApproach} (which is used for the digital control)
affects 
the error signal~(\ref{eqn:ErrorSigY}),
but not (\ref{eqn:SignalPath}).

The NTF (\ref{eqn:scalarinputNTF}) is the starting points of the performance analysis 
in Section~\ref{sec:PerformanceAnalysis}.

\subsection{Bandwidth and the Parameter~$\eta$}
\label{sec:BandwidthEta}

For the following discussion of the parameter $\eta$ 
in (\ref{eqn:EstimFilterDef}) and (\ref{eqn:ADCTransferFunction}),
we restrict ourselves to the scalar-input case,
where the STF and the NTF are given by (\ref{eqn:scalarinputSTF}) and (\ref{eqn:scalarinputNTF}), 
respectively.
In this case, it is easily seen from (\ref{eqn:scalarinputSTF})
that $\eta$ determines the bandwidth
of the estimate (\ref{eqn:DefHatu}).
For example, assuming that $\|\vct{G}(\omega)\|_\infty$ decreases
with $|\omega|$,
the bandwidth is roughly given by $0\leq
|\omega| \leq \omega_\text{crit}$ with $\omega_\text{crit}$ determined by 
\begin{IEEEeqnarray}{rCl} \label{eqn:EtaBandwidth}
\|\vct{G}(\omega_\text{crit})\| &=& \eta.
\end{IEEEeqnarray}

However, the estimate (\ref{eqn:DefHatu})
need not be the final converter output: 
additional filtering is of course possible, 
either in the form of some traditional postfiltering
or via a modification of (\ref{eqn:EstimFilterDef})
as in \cite[Section~IV]{LBWB:ITA2011c} or \cite[Section~VII.A]{BLV:ctfg2013arXiv}

It is also worth noting that
the parameter $\eta$ equals the ratio 
of the STF (\ref{eqn:scalarinputSTF}) 
and 
the NTF at $\omega_\text{crit}$:
\begin{equation} \label{eqn:EtaProxySNR}
\restrict{\frac{\vct{H}(\omega)\vct{G}(\omega)}{\|\vct{H}(\omega)\|}}{\omega=\omega_\text{crit}} = \eta,
\end{equation}
cf.\ \Fig{fig:ADCAmpRp}.

\subsection{Remarks}

We conclude this section with a number of remarks.
First, we note that 
he conversion error (\ref{eqn:CorrErrorSignal}) is not due to the quantizers 
in Figures \ref{fig:GeneralApproach} and \ref{fig:GenSystemSimpleControl}, 
but due to approximating the control-bounded signals $\vct{y}(t)$ by zero
as in (\ref{eqn:SetYZero}).
In particular, 
the quantized signals 
are not used as noisy observations 
of (some filtered version of) $u(t)$, 
but only to determine the digital control. 
In consequence, the quantizer circuits need not be implemented 
with high precision.

Second, 
the STF (\ref{eqn:ADCTransferFunction}) and (\ref{eqn:scalarinputSTF}) 
is an exact continuous-time result.
By contrast, continuous-time $\Delta\Sigma$ modulators are typically 
first designed in discrete time and then converted into continuous time 
using concepts such as direct filter synthesis \cite{OG:2006}.

Finally,
the digital estimation and the transfer function analysis 
of Section~\ref{sec:EstimTFA} work for arbitrary stable 
analog transfer functions $\vct{g}(t)$. 
In fact, stability of the uncontrolled analog system 
has here been assumed only for the sake of the analysis: 
the actual digital filter in Section~\ref{sec:Computation} 
is indifferent to this assumption.
Moreover, 
the details of the digital control (clock frequency, thresholds, etc.)
do not enter the transfer function analysis.
This generality offers design opportunities for the analog system/filter
beyond the limitations 
of conventional $\Delta\Sigma$ modulators.

\section{Conversion Noise Analysis}
\label{sec:PerformanceAnalysis}

In this section, we (again) restrict ourselves to the case 
where $\vct{u}(t)$ 
is scalar (i.e., $k=1$)
and will be denoted by $u(t)$. 
While the analysis in Section~\ref{sec:Theory} was mathematically exact,
we are now prepared to use approximations
similar to those routinely made in the analysis of $\Delta\Sigma$ ADCs.

\begin{figure*}[t]
  \begin{center}
  \setlength{\unitlength}{0.95mm}
  \begin{picture}(188,30)(0,-5)
  \put(0,0){\integratorBox{$\beta_1$}{$-\kappa_1$}}
   \put(5,21.5){\pos{cb}{$u(t)$}}
   \put(22.5,-1.5){\pos{ct}{$s_1(t)$}}
  \put(55,20){\connectionDot}
    \put(55,21.5){\pos{cb}{$x_1(t)$}}
  \put(55,0){\integratorBox{$\beta_2$}{$-\kappa_2$}}
   \put(77.5,-1.5){\pos{ct}{$s_2(t)$}}
  \put(110,20){\connectionDot}
   \put(110,21.5){\pos{cb}{$x_2(t)$}}
  \put(110,20){\line(1,0){5}}
  \put(122.5,20){\cent{$\ldots$}}
  \put(130,0){\integratorBox{$\beta_n$}{$-\kappa_n$}}
   \put(152.5,-1.5){\pos{ct}{$s_n(t)$}}
   \put(180,21.5){\pos{bl}{$x_n(t)$}}
  \end{picture}
  \vspace{3mm}
  \caption{\label{fig:IntegratorChain}%
  Analog part and digital control 
  of the example in Section~\ref{sec:Example}
  for $\rho_1=\ldots=\rho_n=0$.
  }
  \end{center}
\end{figure*}
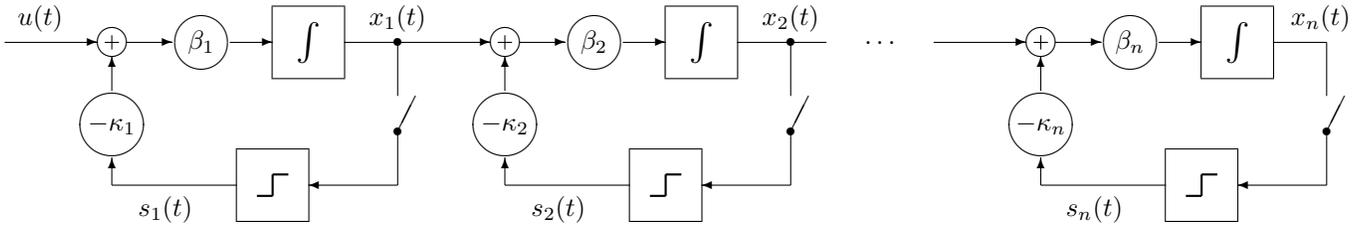

\subsection{SNR and Statistical Noise Model}

Disregarding circuit imperfections 
(which will be addressed in Section~\ref{sec:ThermalNoiseMismatch}),
the quantization performance can be expressed as
the signal-to-noise ratio (SNR) 
\begin{IEEEeqnarray}{rCl} \label{eq:SNR}
  \text{SNR} &\eqdef& \frac{S}{S_\text{N}}
\end{IEEEeqnarray}
where $S$ and $S_\text{N}$ 
are the power of $\hat u(t)$ and the power of 
the conversion error (\ref{eqn:ErrorSigY}),
respectively,
both within some frequency band $\calB$ of interest.

The numerator in (\ref{eq:SNR}) depends, of course, on the input signal. 
A trivial upper bound is $S\leq b_\signame{u}^2$,
and for a full-scale sinusoid, we have
\begin{equation} \label{eqn:SinePower}
S = b_\signame{u}^2/2.
\end{equation}

As for the in-band power $S_\text{N}$ of the conversion error (\ref{eqn:ErrorSigY}),
we begin by writing
\begin{IEEEeqnarray}{rCl} \label{eqn:TotalConvErrorPSD}
\E\!\big[ \epsilon(t)^2 \big] = 
\frac{1}{2\pi}
\int_{-\infty}^{\infty} \vct{H}(\omega) \vct{S}_{\vct{y}\vct{y}^\T}(\omega) \vct{H}(\omega)^\H\, d \omega,
\end{IEEEeqnarray}
where $\vct{y}(t)$ is modeled as a stationary stochastic process
with power spectral density matrix
\begin{IEEEeqnarray}{rCl}
\vct{S}_{\vct{y}\vct{y}^\T}(\omega) &\eqdef& 
\int_{-\infty}^{\infty} \E\!\left[\vct{y}(t + \tau)\vct{y}(t)^{\T}\right] e^{-i \omega \tau}\, d\tau.
\end{IEEEeqnarray}
(These statistical assumptions cannot be literally true,
but they are a useful model.)
Restricting (\ref{eqn:TotalConvErrorPSD}) to the frequency band $\calB$ of interest,
we have
\begin{IEEEeqnarray}{rCl}
S_\text{N} & = &
\frac{1}{2\pi}
\int_{\calB} \vct{H}(\omega) \vct{S}_{\vct{y}\vct{y}^\T}(\omega) \vct{H}(\omega)^\H\, d \omega.
   \label{eqn:GenSN}
\end{IEEEeqnarray}

\subsection{White-Noise Analysis}
\label{sec:GenWhiteNoiseAnalysis}

If $\vct{S}_{\vct{y}\vct{y}^\T}(\omega)$ in (\ref{eqn:GenSN})
is approximated by
\begin{equation} \label{eqn:GenApproxSyy}
\vct{S}_{\vct{y}\vct{y}^\T}(\omega) \approx \sigma_{\vct{y}|\calB}^2 \vct{I}_m,
\end{equation}
we further obtain 
\begin{IEEEeqnarray}{rCl}
S_\text{N} & \approx & 
   \frac{\sigma_{\vct{y}|\calB}^2}{2\pi}
   \int_{\calB} \vct{H}(\omega) \vct{H}(\omega)^\H\, d\omega
   \label{eqn:GenApproxSN} \\
 & = & 
   \frac{\sigma_{\vct{y}|\calB}^2}{2\pi}
   \int_{\calB} \frac{\|\vct{G}(\omega)\|^2}{\big( \|\vct{G}(\omega)\|^2 + \eta^2 \big)^2}\, d\omega
   \IEEEeqnarraynumspace\\
 & \approx & 
   \frac{\sigma_{\vct{y}|\calB}^2}{2\pi}
   \int_{\calB} \frac{1}{\|\vct{G}(\omega)\|^2}\, d\omega,
   \label{eqn:ConversionNoiseWhiteSyy}
\end{IEEEeqnarray}
where the last step is justified 
by $\|\vct{G}(\omega)\| \geq \eta$ for $\omega\in\calB$, 
cf.\ (\ref{eqn:scalarinputSTF}) and Section~\ref{sec:BandwidthEta}.

Note that the approximation (\ref{eqn:GenApproxSyy})
is restricted to $\calB$ 
and is ultimately vindicated 
by the accuracy of~(\ref{eqn:ConversionNoiseWhiteSyy}). 
Using (\ref{eqn:ConversionNoiseWhiteSyy}),
the scale factor $\sigma_{\vct{y}|\calB}^2$
can be determined by simulations.

It is obvious from (\ref{eqn:ConversionNoiseWhiteSyy}) 
that a large SNR (\ref{eq:SNR}) 
requires a large analog amplification, 
i.e., $\|\vct{G}(\omega)\|$ must be large throughout $\calB$.

\section{Example: Chain of Integrators}
\label{sec:Example}

In the following sections, we focus on the specific example shown in \Fig{fig:IntegratorChain}.
(We will return to the general case in Section~\ref{sec:Computation}.)
This example was first presented in \cite{LgW:ITA2015c},
but it is here analyzed much further. 
Other analog circuit topologies with attractive properties 
will be presented elsewhere.

\subsection{Analog Part and Digital Control}

The analog part including the digital control
is shown in \Fig{fig:IntegratorChain}. 
The input signal $u(t)$ is a scalar. 
The state variables $x_1(t)$, \ldots, $x_n(t)$ 
obey the differential equation
\begin{IEEEeqnarray}{rCl}
\frac{d}{dt} x_{\ell}(t) &=& - \rho_\ell x_\ell(t) + \beta_\ell x_{\ell-1} - \kappa_\ell \beta_\ell s_\ell(t)
\end{IEEEeqnarray}
with \mbox{$\rho_\ell\geq 0$}, 
\mbox{$\kappa_\ell \beta_\ell\geq 0$}, 
and with \mbox{$x_{0}(t) \eqdef u(t)$}.
The switches in \Fig{fig:IntegratorChain} represent sample-and-hold circuits 
that are controlled by a digital clock 
with period $T$. 
The threshold elements in \Fig{fig:IntegratorChain} 
produce the control signals 
$s_\ell(t)\in \{ +1, -1 \}$ depending on the sign of $x_\ell(kT)$ 
at sampling time $kT$ immediately preceding $t$.

We will assume $|u(t)| \leq b$,
and the system parameters will be chosen such that 
\begin{IEEEeqnarray}{rCl} \label{eqn:ExampleBoundOnState}
|x_\ell(t)|& \leq &b
\end{IEEEeqnarray}
holds for $\ell=1,\ldots, n$.

The control-bounded signals $y_1(t),\ldots,y_m(t)$ are selected 
from the state variables $x_1(t),\ldots,x_n(t)$ 
(cf.\ \Fig{fig:GenSystemSimpleControl}), 
as will be discussed in Section~\ref{sec:ExampleTransferFunction} below.

\begin{figure}
\begin{center}
\begin{picture}(67.5,30)(0,-25)

\put(0,0){\vector(1,0){13}}   \put(5,1){\pos{cb}{$u(t)$}}
\put(15,0){\plusSign}
\put(17,0){\vector(1,0){6.5}}
\put(27.5,0){\circle{8}}  \put(27.5,0){\cent{$\beta_1$}}
\put(31.5,0){\vector(1,0){6}}
\put(37.5,-5){\framebox(10,10){\Large$\int$}}
\put(47.5,0){\vector(1,0){20}}
\put(55,0){\connectionDot}    \put(55,1.5){\pos{cb}{$x_1(t)$}}
\put(55,0){\line(0,-1){15}}
\put(55,-15){\vector(-1,0){10.5}}
\put(42.5,-15){\plusSign}
 \put(42.5,-25){\vector(0,1){8}}  \put(43.5,-22){\pos{cl}{$e_1(t)$}}
\put(40.5,-15){\vector(-1,0){9}}
\put(27.5,-15){\circle{8}}  \put(27.5,-15){\cent{$-\tilde\kappa_1$}}
\put(23.5,-15){\line(-1,0){8.5}}
\put(15,-15){\vector(0,1){13}}
\end{picture}
\caption{\label{fig:NoMash}%
Conventional view of the first stage in \Fig{fig:IntegratorChain}.}
\end{center}
\end{figure}

\subsection{It's not a MASH Converter}

The system of \Fig{fig:IntegratorChain} has some similarity
with a continuous-time MASH $\Delta\Sigma$ modulator \cite{O:acso2005}.
However, \Fig{fig:IntegratorChain} cannot be handled 
by conventional cancellation schemes. 
To see this, consider \Fig{fig:NoMash},
which shows how the first stage in \Fig{fig:IntegratorChain}
would conventionally be modeled (perhaps with $\tilde\kappa\neq \kappa$),
where $e_1(t)$ is the local quantization error \cite{PST:2017}. 
Since $e_1(t)$ enters the system in exactly the same way as $u(t)$
(except for a scale factor), these two signals 
cannot be separated by any subsequent processing.

By contrast, the digital estimation of Section~\ref{sec:EstimTFA}
cancels the effect of $s_\ell(t)$ on $x_\ell(t)$ in all stages ($\ell=1,\ldots,n$)
and is indifferent to the existence of a conventional cancellation scheme.

\subsection{Conditions Imposed by the Digital Control}
\label{sec:IntChainEx:Control}

The bound (\ref{eqn:ExampleBoundOnState}) 
can be guaranteed by the conditions
\begin{IEEEeqnarray}{rCl} \label{eqn:kappaDef}
|\kappa_\ell| &\geq& b
\end{IEEEeqnarray}
and
\begin{IEEEeqnarray}{rCl} \label{eqn:ControlCondTbeta}
T |\beta_\ell| \big(|\kappa_\ell| + b\big) & \leq & b.
\end{IEEEeqnarray}
With the definition
\begin{equation} \label{eqn:ExGammaEll}
\gamma_\ell \eqdef T |\beta_\ell|,
\end{equation}
(\ref{eqn:ControlCondTbeta}) 
becomes
\begin{equation}
\gamma_\ell \leq \frac{b}{\kappa_\ell + b}
\end{equation}
which implies
$\gamma_\ell \leq 1/2$,
and $\gamma_\ell = 1/2$ is admissible if and only if $\kappa_\ell = b$.
In this case (i.e., if $\kappa_\ell = b$),
the control frequency $1/T$ is admissible if and only if 
\begin{equation} \label{eqn:MinControlFreq}
1/T \geq 2|\beta_\ell|.
\end{equation}

\subsection{Transfer Functions}
\label{sec:ExampleTransferFunction}

As mentioned, the control-bounded signals $y_1(t),\ldots,y_m(t)$ are selected 
from  the state variables $x_1(t),\ldots,x_n(t)$.
An obvious choice is 
$m=n$ and $y_1(t)=x_1(t)$, \ldots, $y_n(t)=x_n(t)$. 
In this case, the ATF 
$\vct{G}(\omega)\eqdef\begin{pmatrix}G_1(\omega) & \dots & G_n(\omega)\end{pmatrix}^\T$ 
of the uncontrolled analog system
(as defined in Section~\ref{sec:Theory})
is given by
\begin{equation} \label{eqn:ExampleTransferFunctionComponent}
G_k(\omega) = \prod_{\ell=1}^k \frac{\beta_\ell}{i\omega+\rho_\ell}
\end{equation}

Another reasonable choice is $m=1$ and $y_1(t)=x_n(t)$ as in \cite{LgW:ITA2015c}.
In this case, the ATF is simply
\begin{equation} \label{eqn:ExampleScalarTransferFunction}
\vct{G}(\omega) = \prod_{\ell=1}^n \frac{\beta_\ell}{i\omega+\rho_\ell}
\end{equation}

We now specialize to the case
where $\beta_1=\ldots=\beta_n=\beta$ 
and
$\rho_1=\ldots=\rho_n=\rho$,
which makes the analysis more transparent.
For $m=1$ as in (\ref{eqn:ExampleScalarTransferFunction}), 
we then have
\begin{equation} \label{eqn:HomIntChainEx:ScalarGSqMag}
\| \vct{G}(\omega) \|^2 
= |G_n(\omega)|^2
= \left( \frac{\beta^2}{\omega^2 + \rho^2} \right)^{\!n}.
\end{equation}
For $m=n$, 
we obtain
\begin{IEEEeqnarray}{rCl} 
\| \vct{G}(\omega) \|^2 
& = & \sum_{k=1}^n 
      |G_k(\omega)|^2 
      \label{eqn:IntChainEx:VctGSqMagSum}\\
& = & \frac{1 - \left( \frac{\omega^2 + \rho^2}{\beta^2} \right)^{\!n}}%
{\left( \frac{\omega^2 + \rho^2}{\beta^2} \right)^{\!n} 
 \left( 1 - \frac{\omega^2 + \rho^2}{\beta^2} \right)}
  \IEEEeqnarraynumspace
  \label{eqn:IntChainEx:VctGSqMag}
\end{IEEEeqnarray}
Note that, for $\omega^2+\rho^2 < \beta^2$,
$|G_n(\omega)|^2$ as in (\ref{eqn:HomIntChainEx:ScalarGSqMag})
is the dominant term in (\ref{eqn:IntChainEx:VctGSqMagSum}).
In consequence, $\vct{G}(\omega)$ as in (\ref{eqn:HomIntChainEx:ScalarGSqMag})
yields almost the same performance as (\ref{eqn:IntChainEx:VctGSqMagSum}).

\begin{figure*}[p]
 \centering
 \begin{minipage}[t]{0.45\linewidth}
 \centering
 \includegraphics[width=0.9\linewidth]{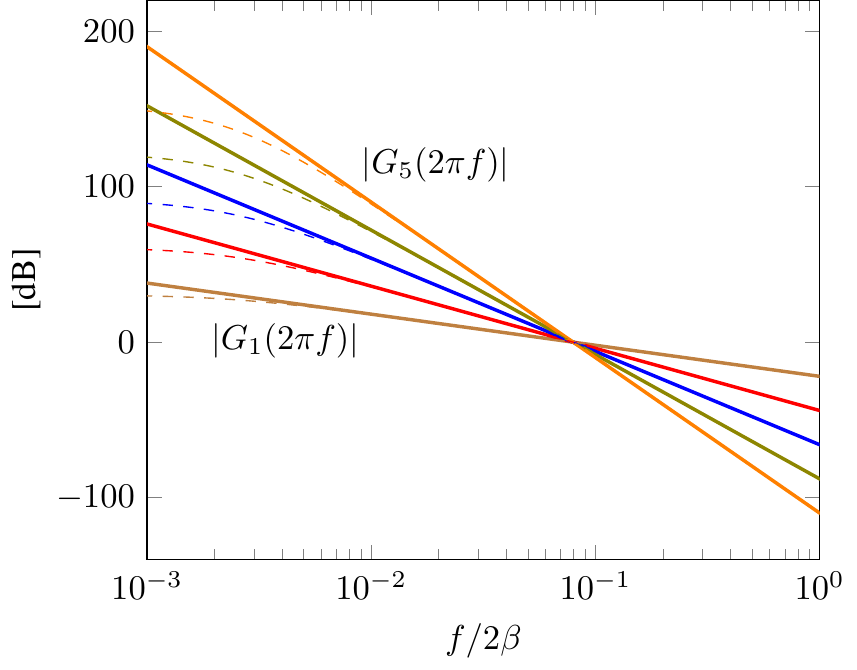}
 \caption{\label{fig:AnalogAmpRp}
 Analog transfer functions (ATF) 
 $|G_1(\omega)|,\,\ldots,\,|G_5(\omega)|$
 of the example in Section~\ref{sec:ExampleTransferFunction},
 with \mbox{$\rho=0$} (solid) 
 and some \mbox{$\rho>0$} (dashed).
 The frequency axis is normalized 
 by the minimum control frequency (\ref{eqn:MinControlFreq}).
 }
 \end{minipage}
 %
 \hspace{3em}
 \begin{minipage}[t]{0.45\linewidth}
 \centering
 \includegraphics[width=0.9\linewidth]{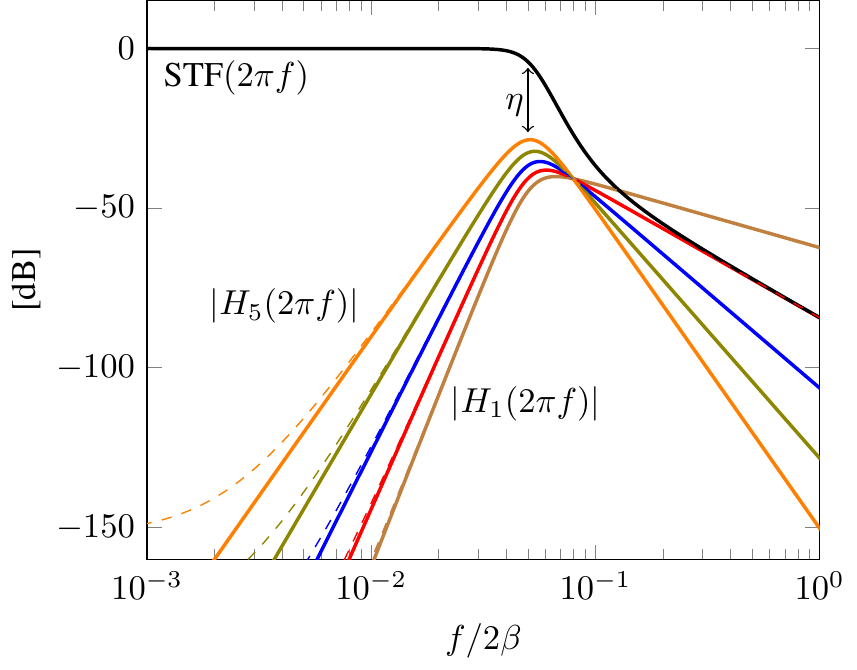}
 \caption{\label{fig:ADCAmpRp}%
 Signal transfer function (STF)
 and noise transfer functions (NTF)
  of the example in Section~\ref{sec:ExampleTransferFunction},
 with \mbox{$\rho=0$} (solid) and some \mbox{$\rho>0$} (dashed).
 }
 \end{minipage}
 %
 \vspace{\dblfloatsep}

 \begin{minipage}[t]{0.45\linewidth}
 \centering
 \includegraphics[width=0.9\linewidth]{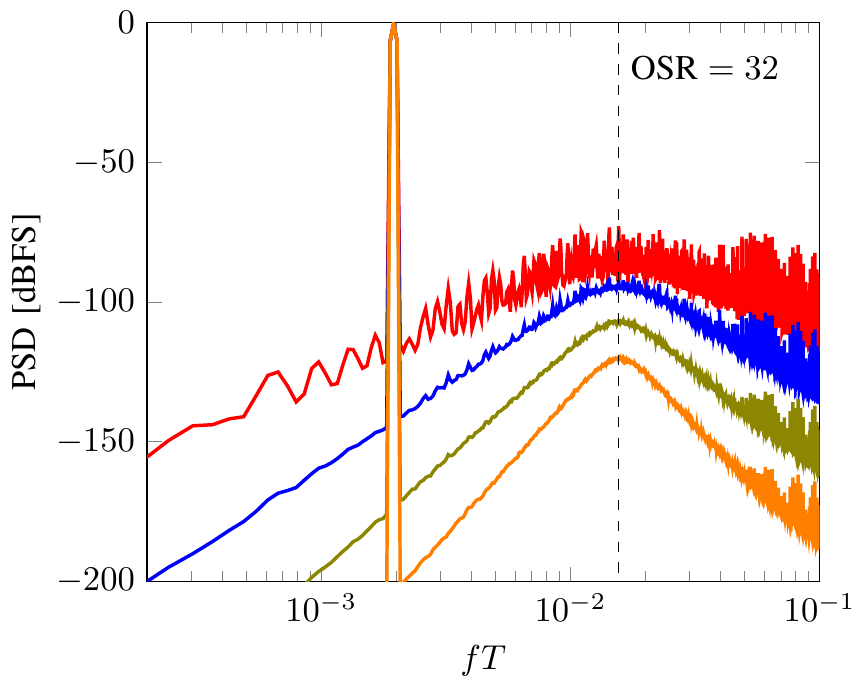}
 \caption{\label{fig:ConvErrPSDSineInput}%
 Simulated power spectral density of the estimate $\hat u(t)$ 
 for the example in 
 Figures \ref{fig:AnalogAmpRp} and~\ref{fig:ADCAmpRp}
 with $n=2, \ldots, 5$ stages 
 and with a full-scale sinusoidal input signal $u(t)$.
 The dashed line indicates the critical frequency defined in (\ref{eqn:EtaBandwidth}).
 }
 \end{minipage}
 %
 \hspace{3em}
 \begin{minipage}[t]{0.45\linewidth}
 \centering
 \includegraphics[width=0.9\linewidth]{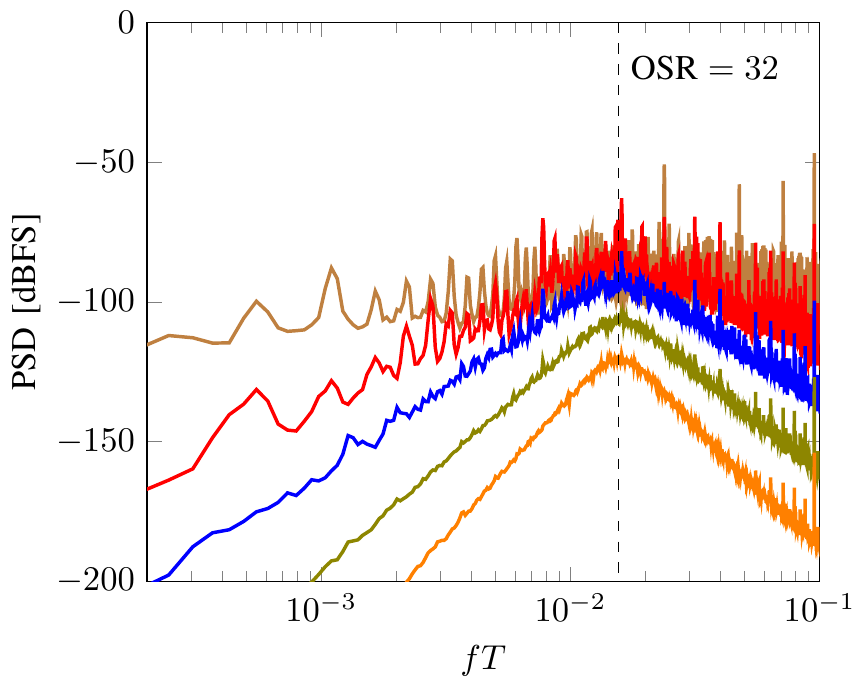}
 \caption{\label{fig:ConvErrPSDNoInput}%
 Same as \Fig{fig:ConvErrPSDSineInput}, but 
 with input signal $u(t)=0$
 and for $n=1, \ldots, 5$.
 }
 \end{minipage}
 \vspace{\dblfloatsep}

 \begin{minipage}[t]{0.45\linewidth}
 \centering
 \includegraphics[width=0.9\linewidth]{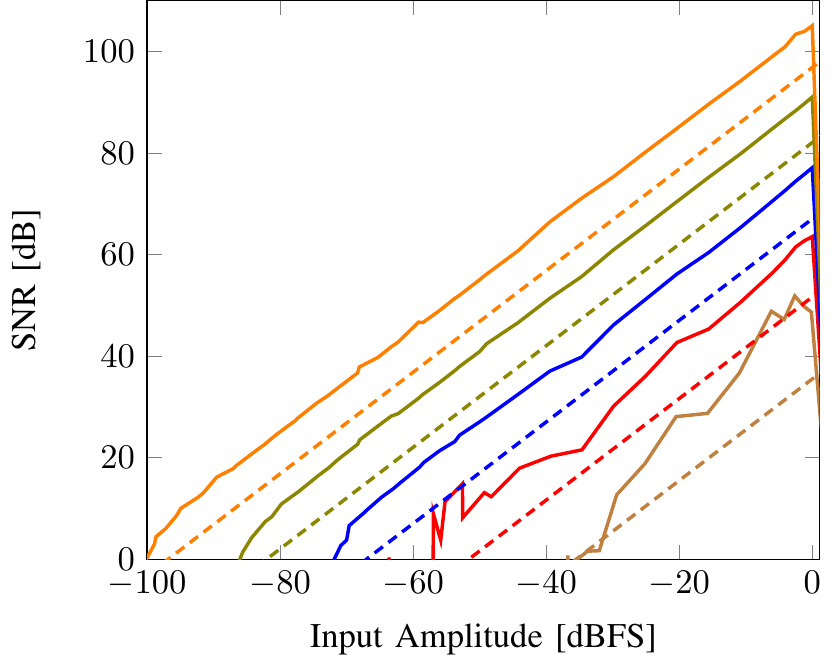}
  \caption{\label{fig:SNRVSINPUTPOWER}%
  The SNR as a function of the input amplitude for $n=1,\ldots,5$ (from right to left). 
  Dashed lines: (\ref{eqn:IntChainEx:WhiteNoiseSNR}) with $\alpha=1$.
  Solid lines: true SNR determined by simulations.}
 \end{minipage}
 \hspace{3em}
 \begin{minipage}[t]{0.45\linewidth}
 \centering
 \includegraphics[width=0.9\linewidth]{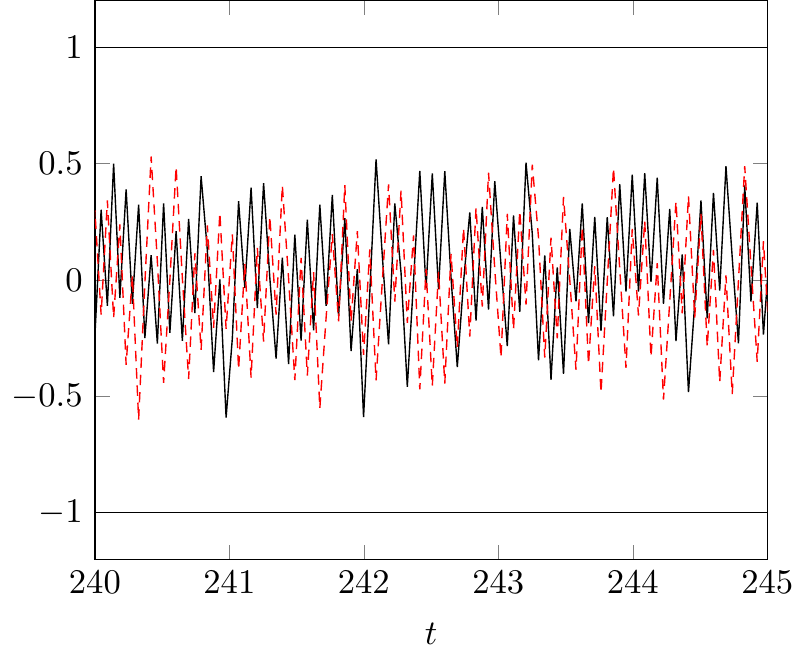}
  \caption{\label{fig:ExSigy}%
  A snapshot of $x_n(t)$ for \mbox{$n=5$} and two different input signals $u(t)$, one of them zero.
  The digital control guarantees \mbox{$|x_n(t)|< 1$}.}
\end{minipage}

\end{figure*}

For illustration, 
the amplitude responses $|G_1(\omega)|$, \ldots, $|G_n(\omega)|$ 
are plotted in \Fig{fig:AnalogAmpRp} for $n=5$,
$\beta=10$, and \mbox{$\rho\in \{ 0, \, 0.03 \beta \}$}.
%
\Fig{fig:ADCAmpRp} shows 
the resulting STF (\ref{eqn:scalarinputSTF}) 
and the components $H_1(\omega)$, \ldots, $H_n(\omega)$
of the NTF (\ref{eqn:scalarinputNTF})
for $m=n$ (i.e., with $\| \vct{G}(\omega) \|$ as in (\ref{eqn:IntChainEx:VctGSqMag}))
and $\eta^2 = 104.3$. 

From now on, we will normally assume $\rho=0$ (i.e., undamped integrators).

\subsection{Bandwidth}
\label{sec:IntChain:Bandwidth}

Using (\ref{eqn:HomIntChainEx:ScalarGSqMag}) (with $\rho=0$), 
the bandwidth $\omega_\text{crit}$ defined by (\ref{eqn:EtaBandwidth})
is easily determined to be
\begin{equation} \label{eqn:IntChainEx:omegacrit}
\omega_\text{crit} = |\beta| / \eta^{\frac{1}{n}}.
\end{equation}
For $\vct G(\omega)$ as in~(\ref{eqn:IntChainEx:VctGSqMag}),
eq.~(\ref{eqn:IntChainEx:omegacrit}) does not strictly hold,
but it is a good proxy for the bandwidth also in this case.

In the following, 
we will use the quantity
\begin{equation} \label{eqn:defOSR}
\text{OSR} \eqdef \frac{1/T}{2 f_\text{crit}}
\end{equation}
with $f_\text{crit} \eqdef \omega_\text{crit}/(2\pi)$,
which may be viewed as an analog of the oversampling ratio of $\Delta\Sigma$ converters.
With (\ref{eqn:IntChainEx:omegacrit}) and with 
\begin{equation} \label{eqn:gammaEqTBeta}
\gamma \eqdef T|\beta|
\end{equation}
as in (\ref{eqn:ExGammaEll}),
we then obtain
\begin{equation} \label{eqn:IntChainEx:OSR2eta}
\eta = {\left( \frac{\gamma}{\pi} \text{OSR} \right)\!}^n.
\end{equation}
Finally, we recall from Section~\ref{sec:IntChainEx:Control}
that stability can be guaranteed if and only if $\gamma\leq 1/2$.

\subsection{Simulation Results}
\label{sec:SimConvNoise}

Figures \ref{fig:ConvErrPSDSineInput} and \ref{fig:ConvErrPSDNoInput}
show the power spectral density (PSD) 
of the digital estimate $\hat u(t)$ 
for the numerical example in Figures \ref{fig:AnalogAmpRp} and~\ref{fig:ADCAmpRp}
with $\rho=0$ and with further details as given below.
In \Fig{fig:ConvErrPSDSineInput}, the input signal $u(t)$
is a full-scale sinusoid;
in \Fig{fig:ConvErrPSDNoInput}, the input signal is $u(t)=0$.
Except for the peak in \Fig{fig:ConvErrPSDSineInput},
both \Fig{fig:ConvErrPSDSineInput} and \Fig{fig:ConvErrPSDNoInput}
thus show the PSD of the conversion error (\ref{eqn:CorrErrorSignal}).

As for the details in these simulations, 
we have $\text{OSR}=32$, $b=1$, $\kappa=1.05$, and $T=1/21.5$,
resulting in $\gamma = 10/21.5$.
The frequency of the sinusoidal input signal is $0.1$~Hz.

The conspicuous fluctuations in the power spectral density 
for \mbox{$n\leq 3$}
can be 
suppressed 
by dithering as described in Section~\ref{sec:ExtraDigitalFeedbackDithering}.

A key point of Figures \ref{fig:ConvErrPSDSineInput} and~\ref{fig:ConvErrPSDNoInput}
is that the PSD of the conversion error (after suitable smoothing by dithering)
appears to be well described by the white-noise analysis of Section~\ref{sec:GenWhiteNoiseAnalysis},
which will be further developed in Section~\ref{sec:IntChainEx:WhiteNoiseAnalysis}.

The resulting SNR (\ref{eq:SNR}), 
as a function of the amplitude of the sinusoidal input signal $u(t)$, 
is shown in \Fig{fig:SNRVSINPUTPOWER}.
Also shown in \Fig{fig:SNRVSINPUTPOWER} 
is an approximate analytical expression 
for the SNR
that will be discussed in Section~\ref{sec:IntChainEx:WhiteNoiseAnalysis}.
Note that the SNR collapses when the input signal amplitude 
exceeds the bound $b$.

\Fig{fig:ExSigy} shows the signal $x_n(t)$
for two different input signals $u(t)$: 
one of the input signals is a sinusoid with frequency $0.1$~Hz and amplitude $1$ 
and the other is $u(t)=0$. 
The point is that the two signals $x_n(t)$ look very much alike: 
like two different realizations of the same stochastic process.
Note also that the digital control, which guarantees \mbox{$x_n(t)<1$}, 
appears to be quite conservative.

\begin{figure*}[]
  \begin{center}
    \begin{minipage}[t]{0.45\linewidth}
      \centering
      \includegraphics[]{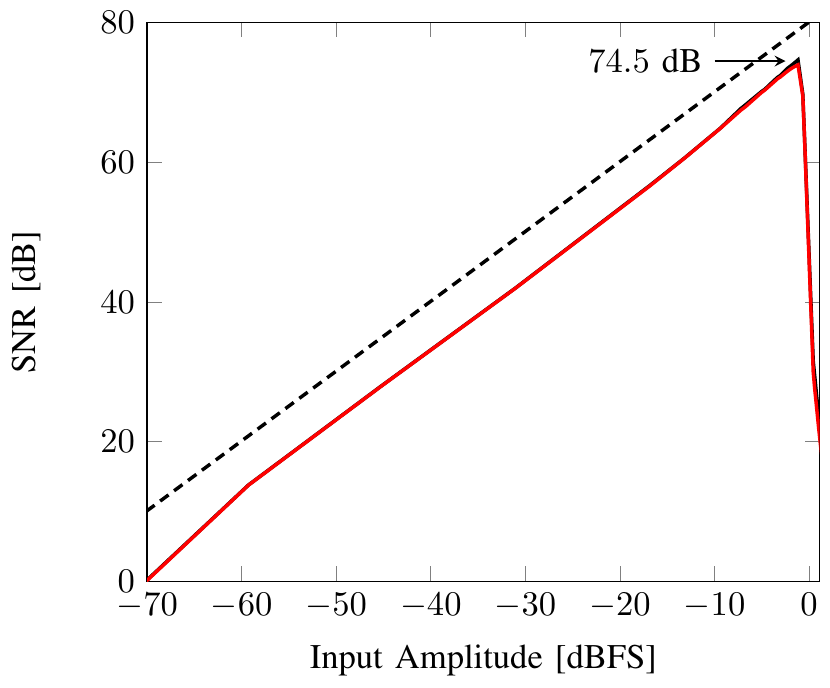}
      \caption{\label{fig:HSNRVSINP} 
      SNR for different input amplitudes of the hardware prototype 
      in Section~\ref{sec:HardwareMeasurements}. 
      Solid black and red (on top of each other): 
      measured SNR and SNDR, respectively.
      Dashed: analytical expression~(\ref{eqn:IntChainEx:WhiteNoiseSNR}) with $\alpha=1$.
      }
    \end{minipage}
    \hspace{3em}
    \begin{minipage}[t]{0.45\linewidth}
      \centering
      \includegraphics[]{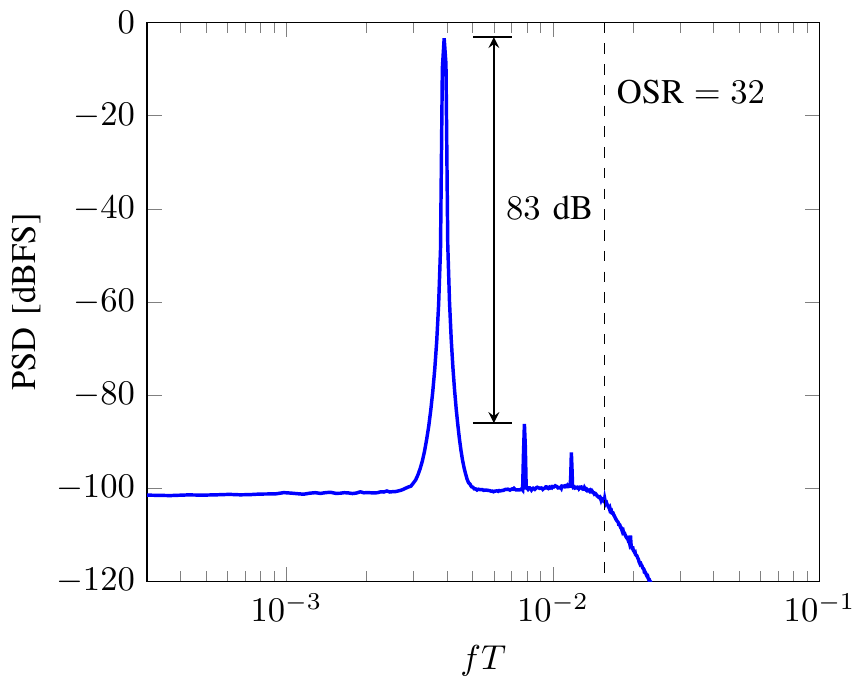}
      \caption{\label{fig:HPSD} Power spectral density of hardware prototype
      corresponding to the largest measured SNR value in Figure
      \ref{fig:HSNRVSINP}.}
    \end{minipage}
  \end{center}
\end{figure*}

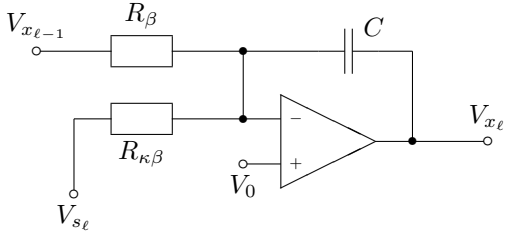
\begin{figure}[]
\begin{center}
\newcommand{\capacitorH}{%
  \begin{picture}(8,4)(0,-2)
  \put(0,0){\line(1,0){3.5}}
  \put(3.5,-3){\line(0,1){6}}
  \put(4.5,-3){\line(0,1){6}}
  \put(4.5,0){\line(1,0){3.5}}
  \end{picture}
}

\begin{picture}(61,29)(7,3)

\put(8,27){\line(1,0){9.5}}
 \put(7.5,27){\circle{1}}      \put(7.5,28.5){\pos{cb}{$V_{x_{\ell-1}}$}}
\put(17.5,25){\framebox(8,4){}}   \put(21.5,30){\pos{cb}{$R_\beta$}}
\put(25.5,27){\line(1,0){19.5}}
\put(35,27){\connectionDot}
 \put(35,27){\line(0,-1){9}}
\put(45,25){\capacitorH}       \put(51,29){\pos{bl}{$C$}}
\put(53,27){\line(1,0){4.5}}
\put(57.5,27){\line(0,-1){12}}

\put(12.5,18){\line(1,0){5}}
 \put(12.5,18){\line(0,-1){9.5}}
 \put(12.5,8){\circle{1}}        \put(12.5,6.5){\pos{ct}{$V_{s_\ell}$}}
\put(17.5,16){\framebox(8,4){}}  \put(21.5,15){\pos{ct}{$R_{\kappa\beta}$}}
\put(25.5,18){\line(1,0){9.5}}

\put(35,18){\connectionDot}
\put(40,18){\line(-1,0){5}}
\put(40,12){\line(-1,0){4.5}}
 \put(35,12){\circle{1}}        \put(35,10.5){\pos{ct}{$V_0$}}
 \put(40,8.75){\line(0,1){12.5}}
 \put(42,18){\pos{c}{\tiny$-$}}
 \put(42,12){\pos{c}{\tiny$+$}}
\put(40,21.25){\line(2,-1){12.5}}
\put(40,8.75){\line(2,1){12.5}}

\put(52.5,15){\line(1,0){14.5}}
 \put(67.5,15){\circle{1}}      \put(67.5,16.5){\pos{cb}{$V_{x_\ell}$}}
\put(57.5,15){\connectionDot}

\end{picture}
\caption{\label{fig:IntegratorCircuit}%
Integrator circuit of the hardware prototype.}
\end{center}
\end{figure}

\subsection{White-Noise Analysis}
\label{sec:IntChainEx:WhiteNoiseAnalysis}

We now explore the white-noise approximation 
(\ref{eqn:GenApproxSyy}) and~(\ref{eqn:ConversionNoiseWhiteSyy})
over the bandwidth 
\begin{equation}
\calB = \{ \omega: |\omega| \leq \omega_\text{crit} \}.
\end{equation}

In the following, we use $\|\vct G(\omega)\|^2$ as in (\ref{eqn:HomIntChainEx:ScalarGSqMag})
(with \mbox{$\rho=0$}),
which also serves as an approximation of (\ref{eqn:IntChainEx:VctGSqMagSum})
provided that $\omega_\text{crit} < |\beta|$.
The integral in (\ref{eqn:ConversionNoiseWhiteSyy}) 
is then easily determined analytically:
\begin{IEEEeqnarray}{rCl}
S_\text{N} & \approx & \frac{\sigma_{\vct{y}|\calB}^2}{2\pi}
    \int_{-\omega_\text{crit}}^{\omega_\text{crit}} \frac{\omega^{2n}}{\beta^{2n}} \, d\omega\\
 & = & 
 \frac{\sigma_{\vct{y}|\calB}^2}{2\pi}
 \cdot \frac{2}{2n+1} \cdot \beta^{-2n} \omega_\text{crit}^{2n+1}.
 \IEEEeqnarraynumspace
\end{IEEEeqnarray}
Using (\ref{eqn:gammaEqTBeta}) and (\ref{eqn:defOSR}), we further obtain
\begin{IEEEeqnarray}{rCl}
S_\text{N} & \approx & 
       \frac{\sigma_{\vct{y}|\calB}^2}{T} \cdot
       \frac{1}{2n+1} \cdot \frac{\pi^{2n}}{\gamma^{2n}} \cdot
       (2f_\text{crit} T)^{2n+1}
       \IEEEeqnarraynumspace\\
 & = & \frac{\sigma_{\vct{y}|\calB}^2}{T} \cdot
       \frac{1}{2n+1} \cdot \frac{\pi^{2n}}{\gamma^{2n}} \cdot
       (\text{OSR})^{-(2n+1)}.
       \IEEEeqnarraynumspace
\end{IEEEeqnarray}

We next approximate the in-band noise power $\sigma_{\vct{y}|\calB}^2$ by
\begin{equation} \label{eqn:IntChainEx:PSDyApprox}
\sigma_{\vct{y}|\calB}^2 \approx \alpha T \frac{(2b)^2}{12}
\end{equation}
with an unknown scale factor $\alpha$.
The factor $b^2$ in (\ref{eqn:IntChainEx:PSDyApprox})
accounts for the bounded amplitude of $y_1(t)$,
cf.\ (\ref{eqn:SinePower}). 
The factor $T$ in (\ref{eqn:IntChainEx:PSDyApprox}) accounts 
for the fact that, for large $n$, 
the signal $y_1(t)=x_n(t)$ originates 
primarily from the control signals $s_1(t),\ldots,s_n(t)$, cf.\ \Fig{fig:ExSigy}. 
Specifically, since $\E[s_\ell(t)^2]=1$ is fixed (for all $\ell=1,\ldots, n$),
the power spectral density%
\footnote{%
Note that we here view $s_1(t),\ldots,s_n(t)$ as stationary stochastic processes.
This cannot be literally true, but it is a useful model. 
Such assumptions/models are standard in the $\Delta\Sigma$ literature.}
of $s_\ell(t)$ scales with $T$. 
Since $x_n(t)$ is essentially created by $s_1(t),\ldots,s_n(t)$
by linear filtering, the power spectral density of $y_1(t)=x_n(t)$ 
scales with $T$ as well.

For large $n$ 
and a sinusoidal input signal $u(t)$ with amplitude $A$, 
the SNR (\ref{eq:SNR}) is thus approximately given by
\begin{equation} \label{eqn:IntChainEx:WhiteNoiseSNR}
\text{SNR} \approx \alpha^{-1} \frac{3A^2}{2b^2} (2n+1) \left( \frac{\gamma}{\pi} \right)^{2n} 
   (\text{OSR})^{2n+1}.
\end{equation}
For the numerical example of Figures 
\ref{fig:AnalogAmpRp}--\ref{fig:ConvErrPSDNoInput},
\Fig{fig:SNRVSINPUTPOWER} shows (\ref{eqn:IntChainEx:WhiteNoiseSNR}) 
for $n>2$ 
to be in good agreement with the actual SNR determined by simulations.

\subsection{Proof of Concept with Hardware Prototype}
\label{sec:HardwareMeasurements}

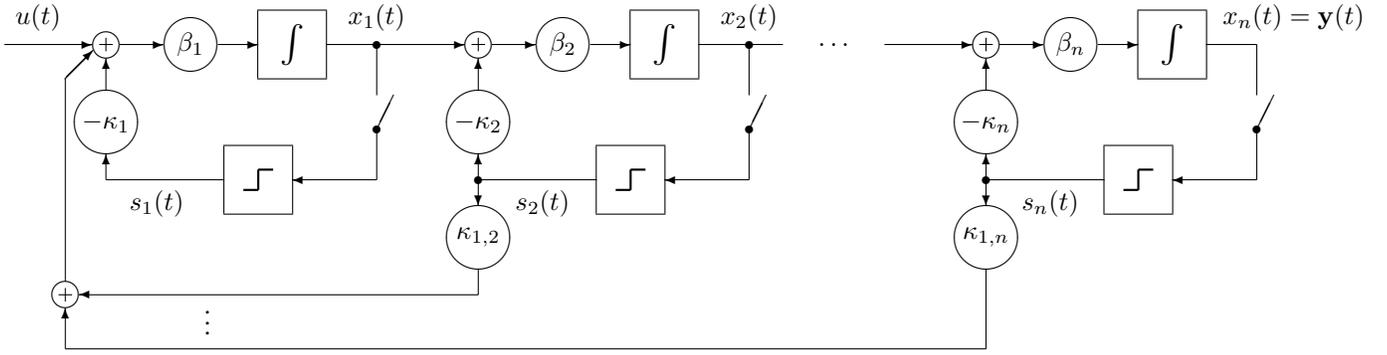
\begin{figure*}[]
  
  \begin{center}
  \setlength{\unitlength}{0.9mm}
  \begin{picture}(200,50)(0,-25)
  %
  \put(0,0){\integratorBox{$\beta_1$}{$-\kappa_1$}}
   \put(5,22){\pos{cb}{$u(t)$}}
   \put(22.5,-1.5){\pos{ct}{$s_1(t)$}}
  \put(55,20){\connectionDot}
    \put(55,22){\pos{cb}{$x_1(t)$}}
  \put(55,0){\integratorBox{$\beta_2$}{$-\kappa_2$}}
   \put(79.5,-1.5){\pos{ct}{$s_2(t)$}}
  \put(110,20){\connectionDot}
   \put(110,22){\pos{cb}{$x_2(t)$}}
  \put(110,20){\line(1,0){5}}
  \put(122.5,20){\cent{$\ldots$}}
  \put(130,0){\integratorBox{$\beta_n$}{$-\kappa_n$}}
   \put(154.5,-1.5){\pos{ct}{$s_n(t)$}}
   \put(180,22){\pos{bl}{$x_n(t) = \vct{y}(t)$}}

  \put(9,-17){\plusSign}
  
  \put(70, 0){\connectionDot}
  \put(70, 0){\vector(0,-1){3.9}}
  \put(70, -8.5){\circle{9} \put(0, 0){\cent{$\kappa_{1,2}$}}}
  \put(70, -13.25){\line(0,-1){3.75}}
  \put(70, -17){\vector(-1,0){59}}
  \put(9, -15){\line(0,1){30}}
  \put(9, 15){\vector(1,1){4.1}}

  \put(30, -20){\cent{$\vdots$}}

  \put(145, 0){\connectionDot}
  \put(145, 0){\vector(0,-1){3.9}}
  \put(145, -8.5){\circle{9} \put(0, 0){\cent{$\kappa_{1,n}$}}}
  \put(145, -13.25){\line(0,-1){11.75}}
  \put(145, -25){\line(-1,0){136}}
  \put(9,-25){\vector(0,1){6}}
  
  \end{picture}
  \vspace{3mm}
  \caption{\label{fig:IntegratorChainWithFeedback}%
  Analog part as in \Fig{fig:IntegratorChain}
  augmented with a little extra digital feedback to prevent limit cycles. 
  }
  \end{center}
\end{figure*}

\begin{figure}
  \centering
  \includegraphics[width=0.9\linewidth]{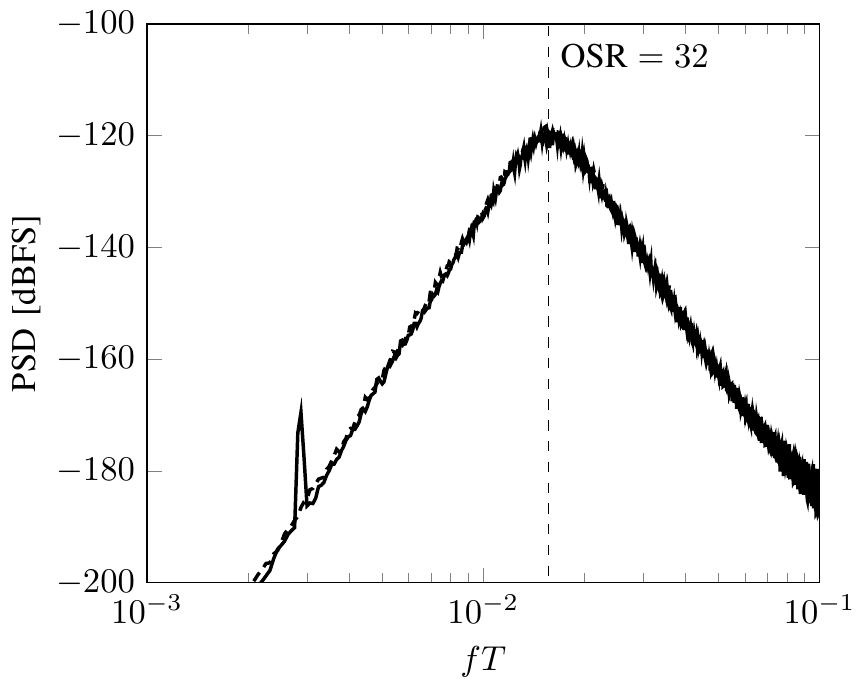}
  \caption{\label{fig:LimitCycles}
  PSD of $\hat u(t)$ 
  as in Figs.\ \ref{fig:GenSystemSimpleControl} and~\ref{fig:IntegratorChain},
  but with a constant input signal (and only for $n=5$).
  Solid: system as in \Fig{fig:IntegratorChain}. 
  Dashed: augmented system as in \Fig{fig:IntegratorChainWithFeedback}.
  }
\end{figure}

\begin{figure}
  \centering
  \includegraphics[]{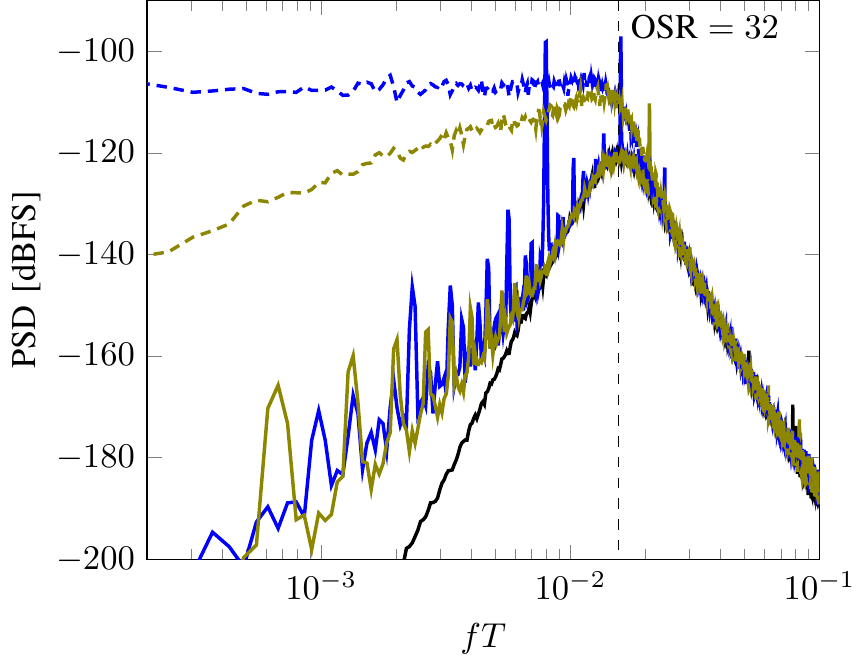}
  \caption{\label{fig:Mismatch} 
  PSD of $\hat u(t)$ 
  for a system with component mismatch and $u(t)=0$.
  Solid: system as in \Fig{fig:IntegratorChain}. 
  Dashed: augmented system as in \Fig{fig:IntegratorChainWithFeedback}. 
  Black (solid only): no mismatch. 
  Blue: $2$\% variation from the nominal value in $\kappa$.
  Green: $2$\% variation in $\beta$.
  The extra feedback coefficients are chosen to have all the same value 
  $\kappa_{1,2}=\ldots=\kappa_{1,n}=\frac{\beta}{n(n-1)}$.
  Note that the dashed lines have no peaks.
}
\end{figure}

Figures \ref{fig:HSNRVSINP} and~\ref{fig:HPSD} show some results with a hardware
prototype
as in \Fig{fig:IntegratorChainWithFeedback}
that was built with
discrete components. 
The only purpose of this prototype was to verify 
the basic functionality of such a converter;
it was not designed to 
excel in terms of speed, accuracy, or power consumption.

Specifically, \Fig{fig:HSNRVSINP} 
shows the measured SNR and SNDR (signal-to-noise-and-distortion ratio)
for a sinusoidal input of frequency $72.4$ Hz, and \Fig{fig:HPSD} shows the PSD for
the measurement corresponding to the largest SNDR in \Fig{fig:HSNRVSINP}. 
We thus have a spurious-free dynamic range (SFDR) of approximately 83 dB 
as well as a SNDR of $74.5$ dB. 

The prototype implements the system of 
\Fig{fig:IntegratorChain}
with $n=5$ nominally identical stages
and with a little additional feedback 
to the first stage as in \Fig{fig:IntegratorChainWithFeedback}
(cf.\ Section~\ref{sec:ExtraDigitalFeedbackDithering}).
The integrators are realized 
with an operational amplifier (AD8615) as shown in \Fig{fig:IntegratorCircuit}.
The control period is chosen as $T=54\mu$s.
The nominal value of the capacitor $C$ is 10\,nF
and the nominal value of both resistors ($R_\beta$ and $R_{\kappa\beta}$)
is 16\,k$\Omega$, resulting in 
$\beta=1/(R_\beta C)=6250/\text{sec}$ and $\kappa=1.25$. For the first stage, the
feedback contributions are
$R_{\kappa_{1,2}\beta}=\ldots=R_{\kappa_{1,5}\beta}=64$\,k$\Omega$. The
operating voltage is 5V, but all signals are confined to the range 0\ldots
2.5\,V; ``zero'' in \Fig{fig:IntegratorChain} translates to $V_0=1.25$\,V. The
resistors and capacitors are standard surface-mount devices with 1\% tolerance;
they were not preselected and their actual values were not measured.

The control signal $s_\ell(t)$ 
(i.e., the voltage $V_{s_\ell}$ in \Fig{fig:IntegratorCircuit})
is generated from $V_{x_\ell}$
using a separate threshold circuit (TLV3201) and a separate analog switch (TS5A9411).
The whole circuitry is realized on a printed circuit board, 
which is piggybacked on an Arduino board.

For the empirical results shown in Figures \ref{fig:HSNRVSINP} and \ref{fig:HPSD},
the digital filter (as described in Section~\ref{sec:Computation}) 
works with nominal values of $\beta$ and $\kappa$; neither 
the hardware prototype nor the digital filter use any calibration
or adjustment for actual (rather than nominal) values.
(See also Section~\ref{sec:ExtraDigitalFeedbackDitheringMismatch}).

The parameter $\eta$ of the digital filter is set 
according to (\ref{eqn:IntChainEx:OSR2eta})
with $\text{OSR}=32$.

\section{Enhancements}
\label{sec:Improvements}

\subsection{Limit Cycles and Dithering by Extra Digital Feedback}
\label{sec:ExtraDigitalFeedbackDithering}

Analog systems as in Figures \ref{fig:GenSystemSimpleControl}
and~\ref{fig:IntegratorChain} may have limit cycles \cite{R:sdto2011},
which lead to distinct peaks
in the power spectrum. 
For example, 
\Fig{fig:LimitCycles} shows the PSD of $\hat u(t)$ 
as in Figures \ref{fig:ConvErrPSDSineInput} and~\ref{fig:ConvErrPSDNoInput},
but with constant input signal $u(t) = 0.003$ (and only for $n=5$):
the conspicuous peak at $fT=0.003$ is due to a limit cycle.

A standard strategy against limit cycles is to use some sort of dithering
\cite{PWG:sqn2007,PG:ldbdith2007,HoKe:mslmash2007,SoPa:spmash2010}. 
Control-bounded converters as in this paper offer convenient ways to do this
without actually adding noise to the estimate $\hat u(t)$.

A first method is to add ``random'' dither to the thresholds for the control signals
$s_1(t),\ldots,s_n(t)$ in Figures \ref{fig:GenSystemSimpleControl}
or~\ref{fig:IntegratorChain}. 
This method does not affect the analysis of Section~\ref{sec:Theory}
and is irrelevant for the digital estimation filter.

A second method, shown in \Fig{fig:IntegratorChainWithFeedback}, 
is to feed small contributions of all the controls 
back to the first stage. This method relies on the effective randomness
of the control signals for large $n$ 
and obviates the need of an extra source of randomness.
Extensive simulations (as exemplified in \Fig{fig:LimitCycles})
have shown this method to be highly effective. 
Note that the augmented system as in \Fig{fig:IntegratorChainWithFeedback}
still fits into the general scheme of \Fig{fig:GeneralApproach}. 
In particular, the extra feedback signals are known to the digital estimation filter,
which can remove their effect on the analog signals. 

When implementing this method, 
it should be noted that the extra feedback 
reduces the allowed amplification for guaranteed
stability of the first stage. However, this reduction is minor 
as the extra feedback can be quite small and yet effective.

\subsection{Helps Against Mismatch, Too}
\label{sec:ExtraDigitalFeedbackDitheringMismatch}

Extra digital feedback as in \Fig{fig:IntegratorChainWithFeedback}
turns out also to significantly mitigate the effects of component mismatch.
For example, the colored solid lines in \Fig{fig:Mismatch} show
the dramatic effect of (simulated) component mismatch on the PSD of $\hat u(t)$
for $u(t)=0$. 
Adding extra digital feedback as 
in \Fig{fig:IntegratorChainWithFeedback} 
raises the floor of the PSD of $\hat u(t)$ 
but removes the detrimental peaks,
resulting in a huge net reduction of the conversion noise.
Extensive simulations 
have shown that this works quite generally.
In particular, 
the good experimental results with the hardware prototype 
shown in Figures \ref{fig:HSNRVSINP} and~\ref{fig:HPSD}
cannot nearly be achieved without this trick.

\subsection{Improving the Scaling in (\ref{eqn:IntChainEx:WhiteNoiseSNR})}
\label{sec:ImprovingGamma}

Except for the factor $\gamma^{2n}$, 
the SNR (\ref{eqn:IntChainEx:WhiteNoiseSNR}) 
scales with $n$ like in a typical $\Delta\Sigma$ converter, 
cf.\ \cite{R:sdto2011}, Eq.\ (4), (7), and~(8).
Recall also (from Sections \ref{sec:IntChainEx:Control} and \ref{sec:IntChain:Bandwidth})
that $\gamma\leq 1/2$ for guaranteed stability.
However, $\gamma$ can be increased beyond $1/2$ in two different ways as follows.

\subsubsection{Venturing Beyond Stability}

As illustrated in \Fig{fig:ExSigy},
the conditions for control as in Section~\ref{sec:IntChainEx:Control} 
may be rather pessimistic. 
Increasing $\gamma_\ell$ beyond 1/2 may thus be ventured for $\ell>1$.
(It should be remembered that conventional high-order $\Delta\Sigma$ converters come without 
stability guarantee.)
However, we have not systematically explored this option.


\subsubsection{Multi-level Quantizers}

Replacing the single-bit quantizers in \Fig{fig:IntegratorChain}
by multi-level quantizers makes the control more effective and 
thereby allows for additional amplification at each stage.
Specifically, using $N$-bit quantizers allows to increase $\gamma$ 
up to
\begin{IEEEeqnarray}{rCl}
  \gamma_\text{max} & = & \frac{1}{2^{(1-N)} + 1}
\end{IEEEeqnarray}
For large $N$, we thus obtain $\gamma \approx 1$.


\section{Thermal Noise and Component Mismatch}
\label{sec:ThermalNoiseMismatch}


The analysis of Sections \ref{sec:PerformanceAnalysis}
and~\ref{sec:IntChainEx:WhiteNoiseAnalysis} 
can be extended to include also thermal noise and component mismatch,
along lines familiar from the analysis of $\Delta\Sigma$ converters.
As in Section~\ref{sec:PerformanceAnalysis}, we restrict
ourselves to the case where $\vct{u}(t) = u(t)$ is scalar (i.e., $k=1$).

\subsection{Thermal Noise}
\label{sec:ThermalNoise}

Let $z(t)$ be a single thermal noise signal 
entering at some point in the analog system
and let $\vct{g}_z(t)$ be the vector of impulse 
responses from this noise source to $y_1(t), \ldots, y_m(t)$.
Thus (\ref{eqn:neteffectOfControl}) is modified into 
\begin{equation}
\vct{y}(t) = (\vct{g} \ast u)(t) + (\vct{g}_z \ast z)(t) - \vct{q}(t).
\end{equation}
In consequence, (\ref{eqn:EstimDecomp}) is modified into
\begin{IEEEeqnarray}{rCl}
\hat u(t) & = & (\vct{h} \ast \vct{q})(t) \\
 & = & (\vct{h} \ast \vct{g} \ast u)(t)
     + (\vct{h} \ast \vct{g}_z \ast z)(t) - (\vct{h} \ast \vct{y})(t),
     \IEEEeqnarraynumspace
\end{IEEEeqnarray}
where the term
\begin{equation} \label{eqn:ThermalNoiseError}
\epsilon_z(t) \eqdef (\vct{h} \ast \vct{g}_z \ast z)(t)
\end{equation}
is the additional error due to $z(t)$. 

Assume now that, within the frequency band $\calB$ of interest, 
$z(t)$ is white with power spectral density 
\begin{equation}
S_z(\omega) = \sigma_{z|\calB}^2.
\end{equation}
The contribution of (\ref{eqn:ThermalNoiseError}) to the noise power (\ref{eqn:GenSN}) 
is then easily determined to be 
\begin{IEEEeqnarray}{rCl}
S_{N,z} & = & \frac{\sigma_{z|\calB}^2}{2\pi} \int_{\calB} \vct{H}(\omega) 
        \vct{G}_z(\omega) \vct{G}_z(\omega)^\H \vct{H}(\omega)^\H\, d\omega 
        \IEEEeqnarraynumspace \label{eqn:ThermalNoiseInBandContrib}\\
 & = & \frac{\sigma_{z|\calB}^2}{2\pi} \int_{\calB} 
       \frac{\| \vct{G}(\omega)^\H \vct{G}_z(\omega) \|^2}{\big(\| \vct{G}(\omega) \|^2 + \eta^2 \big)^2}\, 
       d\omega,
\end{IEEEeqnarray}
where $\vct{G}_z(\omega)$ is the (elementwise) Fourier transform of $\vct{g}_z(t)$.

Finally, the total contribution of multiple such thermal noise sources 
$z_1(t),\, z_2(t),\, \ldots$
to the noise power (\ref{eqn:GenSN}) is simply $S_{N,z_1} + S_{N,z_2} + \ldots$.

\subsection{Mismatch}
\label{sec:Mismatch}

Let $\vct{\tilde g}$, $\vct{\tilde q}$, and $\vct{\tilde h}$ be the nominal 
(i.e., assumed) values of the actual quantities 
$\vct{g}$, $\vct{q}$, and $\vct{h}$, respectively.
We still have
\begin{equation}
\vct{y}(t) = (\vct{g} \ast u)(t) - \vct{q}(t),
\end{equation}
but we now have
\begin{IEEEeqnarray}{rCl}
\hat u(t) & = & (\vct{\tilde h} \ast \vct{\tilde q})(t) \\
  & = &  (\vct{\tilde h} \ast (\vct{\tilde q} - \vct{q}))(t) + (\vct{\tilde h} \ast \vct{q})(t) \\
  & = &  (\vct{\tilde h} \ast (\vct{\tilde q} - \vct{q}))(t)
        + (\vct{\tilde h} \ast \vct{g} \ast u)(t) - (\vct{\tilde h} \ast \vct{y})(t).
       \IEEEeqnarraynumspace
\end{IEEEeqnarray}
The total conversion error can then be written as
\begin{IEEEeqnarray}{rCl}
\epsilon(t) & \eqdef & \hat u(t) - (\vct{\tilde h} \ast \vct{\tilde g} \ast u)(t) \\
  & = & (\vct{\tilde h} \ast (\vct{g} - \vct{\tilde g}) \ast u)(t) 
        + (\vct{\tilde h} \ast (\vct{\tilde q} - \vct{q}))(t)
         \IEEEeqnarraynumspace\nonumber\\
  &  &  {} - (\vct{\tilde h} \ast \vct{y})(t).
            \IEEEeqnarraynumspace
            \label{eqn:MismatchDecomp}
\end{IEEEeqnarray}

The three terms in (\ref{eqn:MismatchDecomp}) are of a very different nature.
The last term, $-(\vct{\tilde h} \ast \vct{y})(t)$, 
is the nominal conversion error (\ref{eqn:ErrorSigY}),
to which the analysis in Section~\ref{sec:PerformanceAnalysis} applies essentially unchanged.
In other words, 
the contribution of this term to the in-band noise power (\ref{eqn:GenSN}) 
is basically unaffected by the mismatch.

The first term in (\ref{eqn:MismatchDecomp}),
\begin{equation} \label{eqn:MismatchErrATF}
\epsilon_\vct{\tilde g}(t) \eqdef 
(\vct{\tilde h} \ast (\vct{g} - \vct{\tilde g}) \ast u)(t),
\end{equation}
accounts for a modification of the STF.
In principle, this term can be neutralized by calibrated postfiltering.  
If this term is considered as noise, its magnitude obviously depends 
on the signal $u(t)$. 
If we assume $u(t)$ to be white noise (within the band $\calB$ of interest),
the contribution of (\ref{eqn:MismatchErrATF}) to the in-band noise power 
can be expressed by an obvious modification of (\ref{eqn:ThermalNoiseInBandContrib}).

The second term in (\ref{eqn:MismatchDecomp}) is more troublesome:
\begin{IEEEeqnarray}{rCl}
\epsilon_\vct{\tilde q}(t) & \eqdef & (\vct{\tilde h} \ast (\vct{\tilde q} - \vct{q}))(t) \\
 & = & \left( \vct{\tilde h} \ast \sum_{\ell=1}^n 
       (\vct{\tilde g}_{q_\ell} - \vct{g}_{q_\ell}) \ast s_\ell \right)\!(t),
      \IEEEeqnarraynumspace \label{eqn:MismatchErrControl}
\end{IEEEeqnarray}
where $\vct{\tilde g}_{q_\ell}$ and $\vct{g}_{q_\ell}$ 
are the nominal and the actual transfer functions, respectively, 
from $s_\ell(t)$ to $y_1(t), \ldots, y_m(t)$.

If we boldly assume $s_1(t), \ldots, s_n(t)$ to be white 
(within the band $\calB$ of interest),
the contribution of (\ref{eqn:MismatchErrControl})
to the in-band noise power can also be expressed 
by an obvious modification of (\ref{eqn:ThermalNoiseInBandContrib}).
However, the white-noise assumption may be too bold, cf.\ \Fig{fig:PSDControlSignals}.
In any case, the power spectral density of $s_1(t), \ldots, s_n(t)$
(for a specific input signal $u(t)$) 
can be determined by simulations, as demonstrated in \Fig{fig:PSDControlSignals}.

\subsection{Application to the Circuit Example}

We now briefly discuss thermal noise and mismatch 
for the circuit example of Section~\ref{sec:Example} 
(Figs.~\ref{fig:IntegratorChain} and~\ref{fig:IntegratorChainWithFeedback})
with integrators as in \Fig{fig:IntegratorCircuit}.
For the sake of clarity, we restrict ourselves to the case $m=1$,
i.e., $y_1(t) = x_n(t)$ is the only control-bounded signal used 
by the digital estimation. 

The transfer function from any noise source at stage $\ell$
to $y_1(t) = x_n(t)$ is
\begin{equation} \label{eqn:NoiseToYComponentTF}
  G_\ell(\omega) = \lambda\left(\frac{\beta}{i\omega}\right)^{n-\ell+1}
\end{equation}
with $\lambda = 1$ for thermal noise
and with 
$\lambda = 1 - \frac{\tilde{\beta}}{\beta}$ or 
$\lambda = 1-\frac{\tilde{\kappa}}{\kappa}$
for mismatch in the resistors in \Fig{fig:IntegratorCircuit},
where $\tilde\beta$ and $\tilde\kappa$ denote the nominal values
while $\beta$ and $\kappa$ denote the actual values.
The product of (\ref{eqn:NoiseToYComponentTF}) and the NTF (\ref{eqn:scalarinputNTF}) 
is the the transfer function from the error source to
the estimate $\hat u(t)$.

In the hardware prototype of Section~\ref{sec:HardwareMeasurements},
all stages (i.e., all integrators) were dimensioned equally. 
Therefore, the prototype will be sensitive primarily to 
errors introduced in the first stage of the chain.

Using the analysis of Section~\ref{sec:ThermalNoise}, 
at room temperature, the thermal noise caused by the two resistors 
in \Fig{fig:IntegratorCircuit} 
should cause a noise floor in $\hat u(t)$
at about $-150$~dB.
It is thus obvious from \Fig{fig:HPSD} 
that thermal noise is not the primary limitation of the prototype.

\begin{figure}[]
  \begin{center}
  \includegraphics[width=0.9\linewidth]{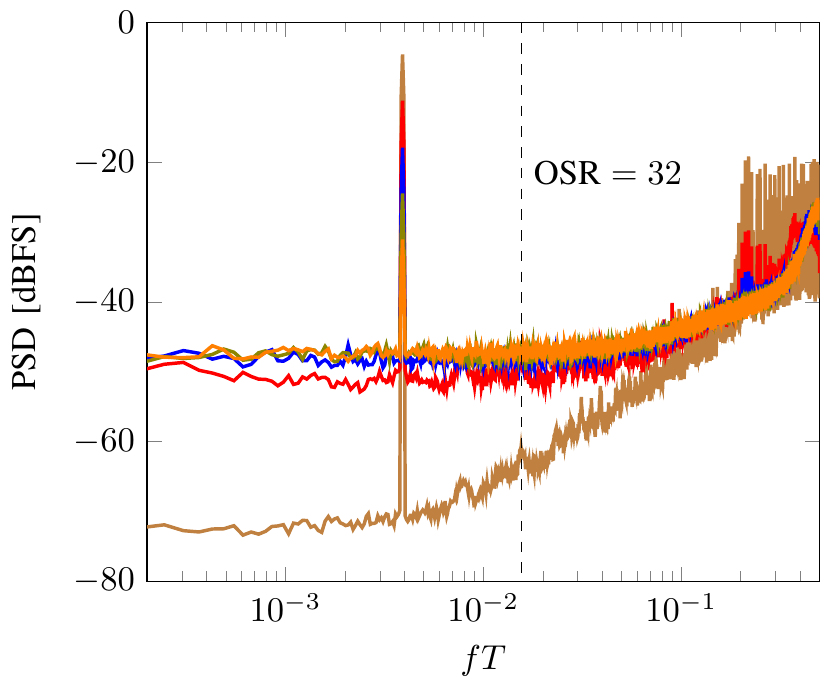}
  \caption{\label{fig:PSDControlSignals} Power spectral density of the control
  signals $s_1,\ldots,s_5$ (brown, red, blue, green, orange) for a system as in
  \Fig{fig:IntegratorChainWithFeedback} excited with a sinusoidal input
  signal with frequency $1/(256T)$ and amplitude $0.7$.}
  \end{center}
\end{figure}

Indeed, the prototype is probably limited primarily by 
component mismatch.
In principle, the analysis of Section~\ref{sec:Mismatch} applies, 
but the PSD of the control signals (as illustrated in \Fig{fig:PSDControlSignals})
defies a simplistic white-noise assumption.
In particular, the PSD of the first-stage control signal $s_1(t)$ 
is very favorably shaped, and this observation seems to be quite stable 
over different experimental scenarios.

\section{Computing $\hat{\vct{u}}(t)$}
\label{sec:Computation}

The job of the digital estimation in \Fig{fig:GeneralApproach} is to compute samples 
of the continuous-time estimate $\hat{\vct{u}}(t)$ defined by (\ref{eqn:DefHatu}) and (\ref{eqn:EstimFilterDef}).
At first sight, this computation looks daunting,
involving not only the continuous-time convolution (\ref{eqn:DefHatu}), but also 
the computation of $\vct{q}(t)$ from the control signals $s_1(t),\ldots,s_n(t)$.

It turns out, however, that samples of $\hat{\vct{u}}(t)$ can be computed quite easily and efficiently 
by the recursions given in Section~\ref{sec:Recursions} below.
A brief derivation of these recursions is given in the Appendix; 
in outline, it involves the following steps.
The starting point is that the filter (\ref{eqn:EstimFilterDef}) 
is formally identical with the optimal filter (the Wiener filter) \cite{KSH:LinEs2000b}\cite{AnMo:of1979b}\
for a certain statistical estimation problem. 
This same statistical estimation problem 
can also be solved by a variation of Kalman smoothing \cite{KSH:LinEs2000b}, 
which leads to recursions based on a state space model of the analog system.
The precise form of the required Kalman smoother is not standard, however,
and it combines input signal estimation as in \cite{BLV:ctfg2013arXiv}
with a limit to continuous-time observations. 

\subsection{State Space Representation of the Analog System}

We will need a state space representation
of the analog system/filter in \Fig{fig:GeneralApproach} of the form
\begin{IEEEeqnarray}{rCl} \label{eqn:StateSpaceModelState}
\frac{d}{dt} \vct{x}(t) &=& \vct{A} \vct{x}(t) + \vct{B} \vct{u}(t) + \vct{\Gamma}\vct{s}(t)
\end{IEEEeqnarray}
and
\begin{IEEEeqnarray}{rCl} \label{eqn:StateSpaceModelOutput}
\vct{y}(t) &=& \vct{C}^\T \vct{x}(t)
\end{IEEEeqnarray}
with state vector $\vct{x}(t)=(x_1(t),\ldots,x_n(t))^\T$, with $\vct{s}(t)
\eqdef (s_1(t),\ldots,s_n(t))^\T$, 
and with real matrices $\vct{A}$, $\vct{B}$,
$\vct{\Gamma}$, and $\vct{C}$ of suitable dimensions. The matrix $e^{\vct{A}}$
will be required to be regular. (As a rule, this regularity condition is
satisfied for ordinary analog filters.)

For the example of Section~\ref{sec:Example} (and \Fig{fig:IntegratorChain}), 
we have
\begin{IEEEeqnarray}{rCl}
\vct{A} & = & \left( \begin{array}{ccccc} 
-\rho_1 & 0 & \hdots & \hdots & 0 \\
\beta_2 & -\rho_2 & 0 & \ddots & \vdots \\
0 & \beta_3 & -\rho_3 & \ddots & \vdots \\
\vdots & \ddots & \ddots & \ddots & 0 \\
0 & \hdots & 0 & \beta_n & -\rho_n
\end{array}\right),
\end{IEEEeqnarray}
$\vct{B} = \big( \beta_1, 0, \ldots, 0\big)^\T$, and
\begin{IEEEeqnarray}{rCl}
\vct{\Gamma} &=& \left( \begin{array}{cccc} 
-\kappa_1\beta_1 & 0 & \hdots & 0 \\
0 & -\kappa_2\beta_2 & \ddots & \vdots \\
\vdots & \ddots & \ddots & 0 \\
0 & \hdots & 0 & -\kappa_n\beta_n
\end{array}\right).
\end{IEEEeqnarray}
If we choose $m=n$ and $y_1(t)=x_1(t)$, \ldots, $y_n(t)=x_n(t)$,
we have $\vct{C}^\T = \vct{I}_{n}$;
if, instead, we choose $m=1$ and $y_1(t)=x_n(t)$, we have
$\vct{C}^\T = (0, \ldots, 0, 1)$.
As stated in Section~\ref{sec:Example}, an obvious
choice for $\rho_1,\ldots,\rho_n$ is $\rho_1=\ldots=\rho_n=0$.

\subsection{Filter Algorithm}
\label{sec:Recursions}

Assume now that we wish to compute $\hat{\vct{u}}(t)$ for $t=t_1,t_2,\ldots$
We will here restrict ourselves to regular sampling%
\footnote{In this section, we use $k$ to index time steps, 
which is unrelated to the dimensionality of $\vct{u}(t)$ as in (\ref{eqn:vctu}).}
with $t_k = k T_u$ such that $T$ 
(the period of the clock in Figures \ref{fig:GeneralApproach} and~\ref{fig:IntegratorChain})
is an integer multiple of $T_u$; 
in other words, we interpolate regularly between the ticks of the clock in \Fig{fig:GeneralApproach}.
(In most practical applications, $T_u=T$ will do.)
Moreover, we focus on the steady-state case $k\gg 1$ where border effects 
can be neglected. 
The algorithm consists of a forward recursion and a backward recursion. 

\begin{trivlist}
\item
\emph{Forward recursion:} for $k=0, 1, 2, \ldots,$ compute the vectors $\vmsgf{m}{k}$
(of the same dimension as $\vct{x}(t)$) by 
\begin{IEEEeqnarray}{rCl} \label{eqn:ForwardRecursion}
\vmsgf{m}{k+1} &\eqdef& \vct{A_f} \vmsgf{m}{k} + \vct{B_f} \vct{s}(t_{k})
\end{IEEEeqnarray}
starting from $\vmsgf{m}{0} \eqdef \vct{0}$.
\end{trivlist}
The required matrices $\vct{A_f}$ and $\vct{B_f}$ will be given
in Section~\ref{sec:AuxMatrices}.

\begin{trivlist}
\item
\emph{Backward recursion:} 
Compute the vectors $\vmsgb{m}{k}$ (of the same dimension as $\vct{x}(t)$) by
\begin{IEEEeqnarray}{rCl} \label{eqn:BackwardRecursion}
\vmsgb{m}{k} &\eqdef& \vct{A_b} \vmsgb{m}{k+1} + \vct{B_b} \vct{s}(t_k)
\end{IEEEeqnarray}
starting from $\vmsgb{m}{N}=\vct{0}$ for some $N>0$,
as well as
\begin{IEEEeqnarray}{rCl} \label{eqn:uhatcomp}
\hat{\vct{u}}(t_k) &=& \vct{W}^\T \! \left(\vmsgb{m}{k} - \vmsgf{m}{k} \right).
  \end{IEEEeqnarray}
\end{trivlist}
The required matrices $\vct{A_b}$ and $\vct{B_b}$ and the matrix $\vct{W}$ 
will be given in Section~\ref{sec:AuxMatrices}.
For the sake of computational efficiency, 
the same starting point $N$ for the backward recursion 
will typically be used to compute (\ref{eqn:uhatcomp})
for a range of consecutive indices $k$.

To be precise, (\ref{eqn:uhatcomp}) agrees with (\ref{eqn:DefHatu}) only 
for $k\gg 0$ and $k\ll N$.
In practice, however, $N-k$ need not be very large for (\ref{eqn:uhatcomp})
to be accurate, 
i.e., only a moderate delay (i.e., latency) is required.

\subsection{FIR Filter and Mixed IIR/FIR Filter Version}

The computation of $\hat{\vct{u}}(t)$ (as described above) can also be organized
as a finite impulse response (FIR) filter or as a mixed IIR/FIR filter. For the
sake of clarity, we consider only the case $T_u=T$, i.e., $t_k = kT$. For the
mixed-filter version, we write (\ref{eqn:uhatcomp}) as
\begin{IEEEeqnarray}{rCl} \label{eqn:uhatcompMixedFilter}
\hat{\vct{u}}(t_k) &=& - \vct{W}^\T \vmsgf{m}{k} 
               + \sum_{\ell=0}^L \vct{\tilde h_\ell\,} \vct{s}(t_{k+\ell})
\end{IEEEeqnarray}
with
\begin{IEEEeqnarray}{rCl}
\vct{\tilde h}_\ell &\eqdef& \vct{W}^\T \vct{A_b^\ell} \vct{B_b}
\end{IEEEeqnarray}
and where the latency parameter $L>0$ replaces $N$.
For the FIR version, the term $\vct{W}^\T \vmsgf{m}{k}$ in (\ref{eqn:uhatcompMixedFilter})
is expanded analogously.

\subsection{Fully Parallel IIR Filter Version}
Equations (\ref{eqn:ForwardRecursion}), (\ref{eqn:BackwardRecursion}) and (\ref{eqn:uhatcomp})
can also be casted as a fully parallel version where
\begin{IEEEeqnarray}{rCl}
  \smsgf{m}{k+1,i} &\eqdef& \smsgf{\lambda}{i} \smsgf{m}{k, i} + \smsgf{f}{i}(\vct{s}(t_{k})) \label{eqn:ParallelForwardMessage} \\
  \smsgb{m}{k,i} &\eqdef& \smsgb{\lambda}{i} \smsgb{m}{k+1, i} + \smsgb{f}{i}(\vct{s}(t_{k})) \label{eqn:ParallelBackwardMessage}
\end{IEEEeqnarray}
and 
\begin{IEEEeqnarray}{rCl} \label{eqn:parallelEstimate}
\hat{u}_j(t_k) &=& \sum_{i=1}^{n} \smsgf{w}{j,i}\smsgf{m}{k,i} + \smsgb{w}{j,i}\smsgb{m}{k,i}
\end{IEEEeqnarray}
Note that (\ref{eqn:ParallelForwardMessage}),
(\ref{eqn:ParallelBackwardMessage}) and (\ref{eqn:parallelEstimate}) are all
scalar expressions. The index $i$ in (\ref{eqn:ParallelForwardMessage})--(\ref{eqn:parallelEstimate}) 
and the index $j$ in (\ref{eqn:parallelEstimate})
refer to the components of the respective vectors.
The coefficients $\smsgf{\lambda}{i}$ and $\smsgb{\lambda}{i}$
are obtained from the eigenvalue decomposition
\begin{IEEEeqnarray}{rCl}
  \vct{A_f} &=& \vct{Q_f} \smsgf{\vct{\Lambda}}{} \vct{Q_f}^{-1} \\
  \vct{A_b} &=& \vct{Q_b} \smsgb{\vct{\Lambda}}{} \vct{Q_b}^{-1}
\end{IEEEeqnarray} where
$\smsgf{\vct{\Lambda}}{} = \diag(\smsgf{\lambda}{1}, \ldots, \smsgf{\lambda}{n})$ and 
$\smsgb{\vct{\Lambda}}{} = \diag(\smsgb{\lambda}{1}, \ldots, \smsgb{\lambda}{n})$ are the
eigenvalues of $\vct{A_f}$ and $\vct{A_b}$ respectively. 
The scalar functions $\smsgf{f}{i}(\cdot)$ and
$\smsgb{f}{i}(\cdot)$ are the $i$-th elements of the vectorized functions
\begin{IEEEeqnarray}{rCl}
  \vmsgf{f}{}(\vct{s}(t_{k})) &\eqdef& \vct{Q_f}^{-1} \vct{B_f}\vct{s}(t_{k}) \label{eqn:ParallelForwardFunction} \\
  \vmsgb{f}{}(\vct{s}(t_{k})) &\eqdef& \vct{Q_b}^{-1} \vct{B_b}\vct{s}(t_{k}), \label{eqn:ParallelBackwardFunction}
\end{IEEEeqnarray}
and $\smsgf{w}{i,j}$ and $\smsgb{w}{i,j}$ are the $(i,j)$-th elements of the
matrices
\begin{IEEEeqnarray}{rCl}
  \vmsgf{W}{} & \eqdef & - \vct{Q_f}^\T \vct{W}\label{eqn:ParallelWForward} \\
  \vmsgb{W}{} & \eqdef & \vct{Q_b}^\T\vct{W}. \label{eqn:ParallelWBackward}
\end{IEEEeqnarray}

Since the components of $\vct{s}(t_k)$ are binary,
the computation of (\ref{eqn:ParallelForwardFunction}) and (\ref{eqn:ParallelBackwardFunction})
from the precomputed columns of $\vct{Q_f}^{-1} \vct{B_f}$
and $\vct{Q_b}^{-1} \vct{B_b}$, respectively,
involves only additions. 
In fact, 
for small $n$, 
(\ref{eqn:ParallelForwardFunction}) and (\ref{eqn:ParallelBackwardFunction})
can be implemented by a lookup table
with $2^n$ entries.

Note that this parallel implementation is computationally very attractive.
In particular, 
if (\ref{eqn:ParallelForwardFunction}) and (\ref{eqn:ParallelBackwardFunction})
can be implemented by lookup tables, 
the computational complexity grows only linearly with $n$.


\subsection{Offline Computations}
\label{sec:AuxMatrices}

We now turn to the matrices $\vct{A_f}, \vct{B_f}, \vct{A_b}, \vct{B_b}$ 
and the matrix $\vct{W}$ in (\ref{eqn:ForwardRecursion})--(\ref{eqn:uhatcomp}),
which can be precomputed.

We first need
the symmetric square matrices $\vmsgf{V}{}$ and $\vmsgb{V}{}$ 
(of the same dimension as $\vct{A}$) as follows. 
The matrix $\vmsgf{V}{}$ is the limit
\begin{IEEEeqnarray}{rCl} \label{eqn:DefVf}
\vmsgf{V}{} &\eqdef& \lim_{\tau\rightarrow 0} \lim_{\ell\rightarrow \infty} \msgf{\vct{V}}{\ell}
\end{IEEEeqnarray}
of the iteration
\begin{IEEEeqnarray}{rCl}
\msgf{\vct{V}}{\ell+1}
& \eqdef & \msgf{\vct{V}}{\ell} + \tau \Bigg(
  \vct{A} \msgf{\vct{V}}{\ell} + (\vct{A} \msgf{\vct{V}}{\ell})^\T 
  \nonumber\\
&& 
  \mbox{\hspace{2em}} + \vct{B}\vct{B}^\T
  - \frac{1}{\eta^2} \msgf{\vct{V}}{\ell}\vct{C}\vct{C}^\T \msgf{\vct{V}}{\ell}
\Bigg);
\IEEEeqnarraynumspace
\label{eqn:DefVfIter}
\end{IEEEeqnarray}
equivalently, $\vmsgf{V}{}$ is the solution of 
the continuous-time algebraic Riccati equation
\begin{IEEEeqnarray}{C} \label{eqn:ForwRiccati}
\vct{A} \msgf{\vct{V}}{} + (\vct{A} \msgf{\vct{V}}{})^\T 
+ \vct{B}\vct{B}^\T
  - \frac{1}{\eta^2} \msgf{\vct{V}}{}\vct{C}\vct{C}^\T \msgf{\vct{V}}{}
  = \vct{0}.
  \IEEEeqnarraynumspace
\end{IEEEeqnarray}

The matrix $\vmsgb{V}{}$ is defined almost identically, but with a sign change
in $\vct{A}$, i.e., $\vmsgb{V}{}$ is the solution 
of the continuous-time algebraic Riccati equation
\begin{IEEEeqnarray}{C} \label{eqn:BackwRiccati}
\vct{A} \msgb{\vct{V}}{} + (\vct{A} \msgb{\vct{V}}{})^\T 
- \vct{B}\vct{B}^\T
  + \frac{1}{\eta^2} \msgb{\vct{V}}{}\vct{C}\vct{C}^\T \msgb{\vct{V}}{}
  = \vct{0}.
  \IEEEeqnarraynumspace
\end{IEEEeqnarray}

The matrix $\vct{W}$ in (\ref{eqn:uhatcomp}) is then obtained by solving
the linear equation
\begin{IEEEeqnarray}{rCl} \label{eqn:wEquation}
( \vmsgf{V}{} + \vmsgb{V}{} ) \vct{W} &=& \vct{B}
\end{IEEEeqnarray}
for $\vct{W}$.

The matrix $\vct{A_f}$ in (\ref{eqn:ForwardRecursion}) is given by
\begin{IEEEeqnarray}{rCl} \label{eqn:Af}
\vct{A_f} &\eqdef& e^{(\vct{A} - \vmsgf{V}{}\vct{C}\vct{C}^\T/\eta^2)T_u}
\end{IEEEeqnarray}
and the matrix $\vct{A_b}$ in (\ref{eqn:BackwardRecursion}) is 
\begin{IEEEeqnarray}{rCl}
\vct{A_b} &\eqdef& e^{-(\vct{A} + \vmsgb{V}{}\vct{C}\vct{C}^\T/\eta^2)T_u}.
\end{IEEEeqnarray}
Finally, the matrix $\vct{B_f}$ in (\ref{eqn:ForwardRecursion}) is 
\begin{IEEEeqnarray}{rCl} \label{eqn:Bf}
\vct{B_f} &\eqdef&
\int_{0}^{T_u} \! e^{(\vct{A} - \vmsgf{V}{}\vct{C}\vct{C}^\T/\eta^2)(T_u-t)} \vct{\Gamma}\, dt 
\end{IEEEeqnarray}
and the matrix $\vct{B_b}$ in (\ref{eqn:BackwardRecursion}) is
\begin{IEEEeqnarray}{rCl}
\vct{B_b} &\eqdef& 
-\int_{0}^{T_u} \! e^{-(\vct{A} + \vmsgb{V}{}\vct{C}\vct{C}^\T/\eta^2)(T_u-t)} \vct{\Gamma}\, dt. 
\end{IEEEeqnarray}

Note that the only free parameter of the digital filter 
is $\eta^2$ as in (\ref{eqn:EstimFilterDef}).

Care must be taken that the quantities of this section
are computed with sufficient numerical precision, 
and the matrices $\vmsgf{V}{}$ and $\vmsgb{V}{}$ should be exactly symmetric. 

For the example of Section~\ref{sec:Example} (and \Fig{fig:IntegratorChain})
with $n=2$ and $\rho=0$, the quantities in (\ref{eqn:wEquation}) turn out to be
\begin{IEEEeqnarray}{rCl}
\vmsgf{V}{} &=& \left(\begin{array}{cc}
   \beta\sqrt{2\eta} & \beta\eta \\
    \beta\eta & \beta\eta\sqrt{2\eta}
  \end{array} \right),
\end{IEEEeqnarray}
\begin{IEEEeqnarray}{rCl}
\vmsgb{V}{} &=& \left(\begin{array}{cc}
   \beta\sqrt{2\eta} & -\beta\eta \\
    -\beta\eta & \beta\eta\sqrt{2\eta}
  \end{array} \right),
\end{IEEEeqnarray}
and 
$\vct{W} = \frac{1}{2\sqrt{2\eta}}(1, 0)^\T$, 
which may be a useful test case for numerical computations.

\section{Conclusion}
\label{sec:Concl}

We have developed the principles, and discussed many details, 
of control-bounded analog-to-digital conversion. 
Such converters have many commonalities with $\Delta \Sigma$
converters, but they can employ analog systems/filters 
for which no traditional cancellation scheme exists.
While we gave an example of such an ADC (in Section~\ref{sec:Example}), 
it should be clear that many other circuit topologies are possible.
(Some such topologies with attractive properties will be presented elsewhere.)
The present paper
provides sufficient information for analog designers 
to experiment with such ADCs.

\appendix[Brief Derivation of the Digital Filter Algorithm]

\begin{figure*}[]
\begin{center}
\begin{picture}(164,78)(-2,7)

\put(0,50){\cent{$\ldots$}}
\put(5,50){\vector(1,0){20}}  \put(7,51){\pos{bl}{$\vct{X}(t_{k-1})$}}
\put(25,45){\framebox(10,10){$e^{\vct{A}\Delta}$}}
\put(35,50){\vector(1,0){12.5}}
\put(47.5,47.5){\framebox(5,5){$+$}}

\put(47.5,80){\framebox(5,5){}}  \put(46.25,85){\pos{tr}{$\calN(\vct{0},\frac{\sigma_U^2}{\Delta} \vct{I}_K)$}}
\put(50,80){\vector(0,-1){10}}   \put(51,75){\pos{cl}{$\tilde{\vct{U}}(t_k,\Delta)$}}
\put(46,62){\framebox(8,8){$\vct{B}\Delta$}}
\put(50,62){\vector(0,-1){9.5}}

\put(52.5,50){\vector(1,0){12.5}}
\put(65,47.5){\framebox(5,5){$=$}}
\put(70,50){\vector(1,0){20}}    \put(80,51){\pos{bc}{$\vct{X}(t_k)$}}

\put(67.5,47.5){\vector(0,-1){9.5}}
\put(63.5,30){\framebox(8,8){$\vct{C}^\T$}}
\put(67.5,30){\vector(0,-1){7.5}}
\put(65,17.5){\framebox(5,5){$+$}}

\put(47.5,17.5){\framebox(5,5){}} \put(48,23.75){\pos{bc}{$\calN(\vct{0},\frac{\sigma_Z^2}{\Delta} \vct{I}_m)$}}
\put(52.5,20){\vector(1,0){12.5}}

\put(67.5,17.5){\vector(0,-1){7.5}}
\put(67.5,9){\knownBox}  \put(69.5,9){\pos{cl}{$\vct{q}(t_k)$}}

\put(90,45){\framebox(10,10){$e^{\vct{A}\Delta}$}}
\put(100,50){\vector(1,0){12.5}}
\put(112.5,47.5){\framebox(5,5){$+$}}

\put(112.5,80){\framebox(5,5){}}  \put(111.25,85){\pos{tr}{$\calN(\vct{0},\frac{\sigma_U^2}{\Delta} \vct{I}_K)$}}
\put(115,80){\vector(0,-1){10}}   \put(116,75){\pos{cl}{$\tilde{\vct{U}}(t_{k+1},\Delta)$}}
\put(111,62){\framebox(8,8){$\vct{B}\Delta$}}
\put(115,62){\vector(0,-1){9.5}}

\put(117.5,50){\vector(1,0){12.5}}
\put(130,47.5){\framebox(5,5){$=$}}
\put(135,50){\vector(1,0){20}}    \put(145,51){\pos{bc}{$\vct{X}(t_{k+1})$}}
\put(160,50){\cent{$\ldots$}}

\put(132.5,47.5){\vector(0,-1){9.5}}
\put(128.5,30){\framebox(8,8){$\vct{C}^\T$}}
\put(132.5,30){\vector(0,-1){7.5}}
\put(130,17.5){\framebox(5,5){$+$}}

\put(112.5,17.5){\framebox(5,5){}} \put(113,23.75){\pos{bc}{$\calN(\vct{0},\frac{\sigma_Z^2}{\Delta}\vct{I}_m)$}}
\put(117.5,20){\vector(1,0){12.5}}

\put(132.5,17.5){\vector(0,-1){7.5}}
\put(132.5,9){\knownBox}  \put(134.5,9){\pos{cl}{$\vct{q}(t_{k+1})$}}

\end{picture}
\caption{\label{fig:FactorGraphUncontrolled}%
Two sections of the factor graph of the (uncontrolled) state space model. 
The total factor graph consists of many such sections; perhaps with initial and
final conditions, which we can ignore in this paper. A box labeled
``$\calN(\vct{m},\vct{\Sigma})$'' represents a multivariate Gaussian density
with mean vector $\vct{m}$ and covariance matrix $\vct{\Sigma}$, $\vct{0}$
refers to an all zero vector of appropriate dimensions, and a small filled
box represents a known quantity; all other boxes represent linear equations.
This factor graph representation is exact only in the limit $\Delta=t_k-t_{k-1}
\rightarrow 0$.}
\end{center}
\end{figure*}
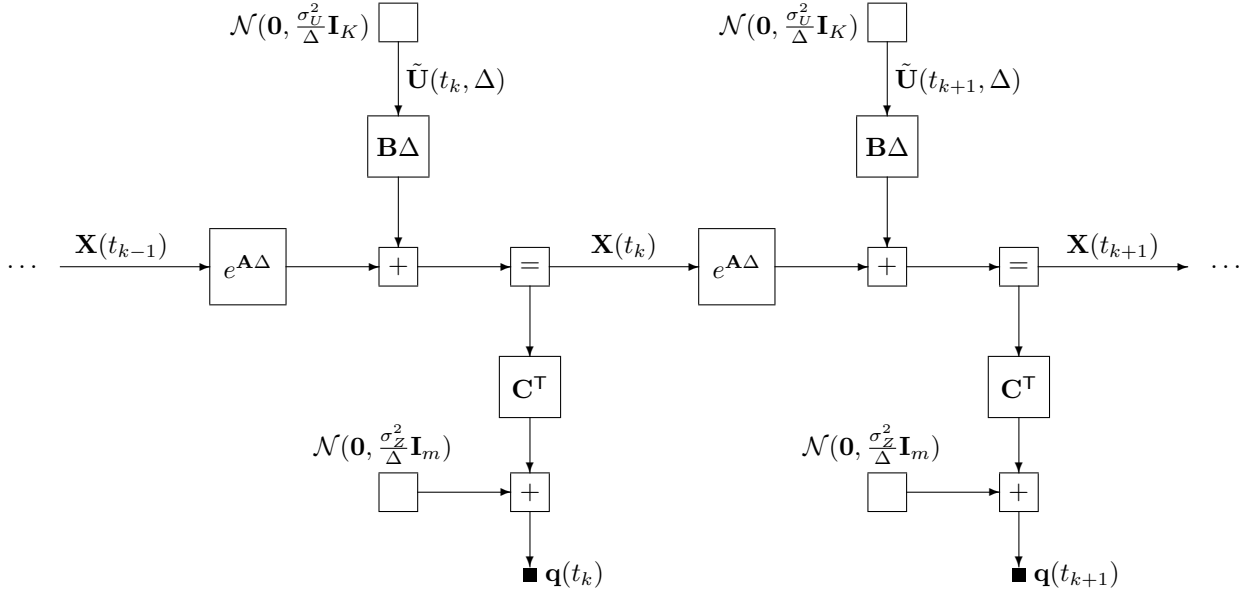

\begin{figure}[]
\begin{center}
\begin{picture}(85,77)(0,8)

\put(0,50){\vector(1,0){15}}    \put(6.5,51){\pos{cb}{$\vct{X}(t_{k-1})$}}
\put(15,45){\framebox(10,10){$e^{\vct{A}\Delta }$}}
\put(25,50){\vector(1,0){10}}
\put(35,47.5){\framebox(5,5){$+$}}

\put(37.5,80){\knownBox}    \put(35.5,80){\pos{cr}{$\vct{s}(t_{k-1})$}}
\put(37.5,80){\vector(0,-1){10}}
\put(33.5,62){\framebox(8,8){$\vct{\Gamma}\Delta$}}
\put(37.5,62){\vector(0,-1){9.5}}

\put(40,50){\vector(1,0){10}}
\put(50,47.5){\framebox(5,5){$+$}}

\put(50,80){\framebox(5,5){}}  \put(56.25,85){\pos{tl}{$\calN(\vct{0},\frac{\sigma_U^2}{\Delta} \vct{I}_{K})$}}
\put(52.5,80){\vector(0,-1){10}}   \put(53.5,75){\pos{cl}{$\tilde{\vct{U}}(t_k,\Delta)$}}
\put(48.5,62){\framebox(8,8){$\vct{B}\Delta$}}
\put(52.5,62){\vector(0,-1){9.5}}

\put(55,50){\vector(1,0){10}}    \put(60,51){\pos{cb}{\small$\vct{X}(t_k^-)$}}
\put(65,47.5){\framebox(5,5){$=$}}
\put(70,50){\vector(1,0){15}}    \put(77.5,51){\pos{bc}{$\vct{X}(t_k)$}}

\put(67.5,47.5){\vector(0,-1){9.5}}
\put(63.5,30){\framebox(8,8){$\vct{C}^\T$}}
\put(67.5,30){\vector(0,-1){7.5}}
\put(65,17.5){\framebox(5,5){$+$}}

\put(47.5,17.5){\framebox(5,5){}} \put(48,23.75){\pos{bc}{$\calN(\vct{0},\frac{\sigma_Z^2}{\Delta}\vct{I}_{m})$}}
\put(52.5,20){\vector(1,0){12.5}}

\put(67.5,17.5){\vector(0,-1){7.5}}
\put(67.5,9){\knownBox}  \put(69.5,9){\pos{cl}{$\vct{0}$}}

\end{picture}
\caption{\label{fig:FactorGraph}%
One section of the factor graph of the state space model with plugged-in digital control
signals $\vct{s}(t)$. The total factor graph consists of many such sections.
The representation is exact only in the limit $\Delta = t_k-t_{k-1} \rightarrow 0$,
where $e^{\vct{A}\Delta}\rightarrow \vct{I}_{n} + \vct{A}\Delta$.}
\end{center}
\end{figure}
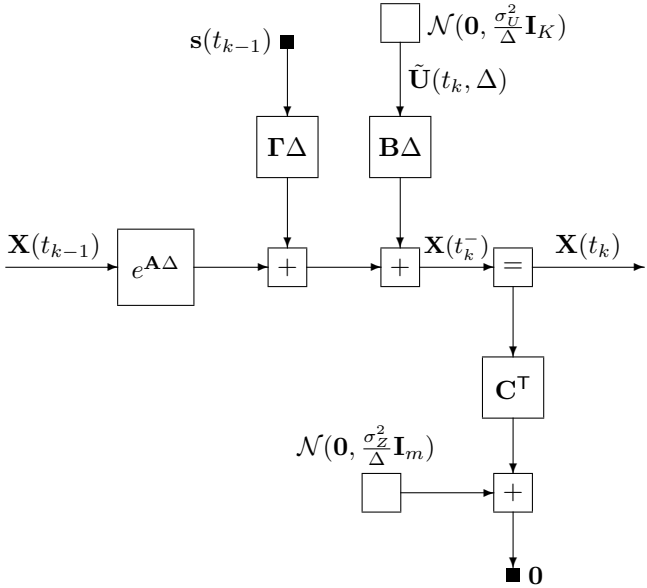

In this appendix, 
we give a condensed derivation
of the algorithm of Section~\ref{sec:Computation}. 
(A detailed development of all the required
background is beyond the scope of this paper.)

We first observe that the filter (\ref{eqn:EstimFilterDef}) 
is formally a multivariate extension of the continuous-time Wiener filter \cite{AnMo:of1979b}
that estimates a multivariate zero-mean white Gaussian noise ``signal'' $\vct{U}(t)$ 
from the signal 
\begin{IEEEeqnarray}{rCl} \label{eqn:WienerObservation}
\tilde{\vct{Y}}(t) &\eqdef& (\vct{g} \ast \vct{U})(t) + \vct{Z}(t),
\end{IEEEeqnarray}
where $\vct{Z}(t)$ is $m$-dimensional zero-mean white Gaussian noise
that is independent of $\vct{U}(t)$. 
In this statistical model, the average
\begin{IEEEeqnarray}{rCl} \label{eqn:DefLocalAverageUt}
\tilde{\vct{U}}(t,\Delta) &\eqdef& \frac{1}{\Delta} \int_{t-\Delta}^t \vct{U}(\tau)\, d\tau
\end{IEEEeqnarray}
(for $\Delta>0$) 
is a $K$-dimensional%
\footnote{In this appendix, we use $K$, rather than $k$ as in (\ref{eqn:vctu}),
to denote the number of input signals.}
zero-mean Gaussian random variable 
with covariance matrix
$\frac{\sigma_U^2}{\Delta} \vct{I}_{K}$. 
The covariance matrix $\frac{\sigma_Z^2}{\Delta}\vct{I}_{m}$ 
of $\vct{Z}(t)$ is defined analogously.

By ``estimating $\vct{U}(t)$'', we really mean to estimate 
the random variable(s) (\ref{eqn:DefLocalAverageUt}) 
for any fixed $t$, and then taking the limit $\Delta\rightarrow 0$ \cite{BrLg:sise2014c}.
In this setting, the MAP estimate, the MMSE estimate, and the LMMSE estimate 
agree and equal the mean of the posterior distribution of 
$\tilde{\vct{U}}(t,\Delta)$ conditioned on the observation of $\tilde{\vct{Y}}(t)$. 
The Wiener filter computes this estimate (for $\Delta\rightarrow 0$) as 
\begin{IEEEeqnarray}{rCl}
\hat{\vct{U}}(t) &=& (\vct{h} \ast \tilde{\vct{Y}})(t)
\end{IEEEeqnarray}
where the Fourier transform of $\vct{h}(t)$ is (\ref{eqn:EstimFilterDef}) with 
\begin{IEEEeqnarray}{rCl}
\eta^2 &=& \sigma_Z^2 / \sigma_U^2.
\end{IEEEeqnarray}

Applying this Wiener filter to the signal $\vct{q}(t)$ as in (\ref{eqn:DefHatu})
means that we solve the statistical estimation problem 
for the observation $\tilde{\vct{Y}}(t) = \vct{q}(t)$.

The same statistical estimation problem can also be solved by 
a variation of Kalman smoothing. 
In constrast to the Wiener filter, the Kalman approach is based 
on the state space equations (\ref{eqn:StateSpaceModelState}) and (\ref{eqn:StateSpaceModelOutput}),
which leads to recursive estimation algorithms. 
We will use a discrete-time approximation of the state space model 
with discrete times%
\footnote{The discrete times $t_1, t_2, \ldots$ in this appendix 
(with $t_k - t_{k-1}=\Delta \rightarrow 0$) are unrelated 
to the discrete time steps in Section~\ref{sec:Computation}.} 
$t_1, t_2, \ldots$ and fixed $t_k - t_{k-1} = \Delta>0$;
our continuous-time results will then be obtained by taking the limit $\Delta\rightarrow 0$.

From now on, we will use factor graphs as in \cite{LDHKLK:fgsp2007}, 
which allow to compose recursive estimation algorithms 
from lookup tables of ``local'' computations.
A factor graph of (the discrete-time approximation of) 
our statistical model in state space form is shown in \Fig{fig:FactorGraphUncontrolled}. 
Note that \Fig{fig:FactorGraphUncontrolled} represents the uncontrolled analog system 
with the observations $\tilde{\vct{Y}}(t_k) = \vct{q}(t_k)$.

Now we plug in the (known and piecewise constant) control signals 
$\vct{s}(t) = (s_1(t),\ldots, s_n(t))$ into the state space model. 
We thus obtain the factor graph of \Fig{fig:FactorGraph}, where
all the observed signals are now zero,
cf.\ (\ref{eqn:SetYZero}).
This second factor graph is easy to work with 
and then to take the limit $\Delta\rightarrow 0$ to continuous time. 

Using the notation of \cite{LDHKLK:fgsp2007}, we now consider 
the quantities $\vmsgf{m}{\vct{X}(t)}$ and $\vmsgf{V}{\vct{X}(t)}$
as well as $\vmsgb{m}{\vct{X}(t)}$ and $\vmsgb{V}{\vct{X}(t)}$.
The former denote the mean vector and the covariance matrix,
respectively, of the forward sum-product message, 
which equals the Gaussian probability density of the time-$t$ state 
$\vct{X}(t)$ given past observations (up to a scale factor);
the latter denote the mean vector and the covariance matrix,
respectively, of the backward sum-product message, 
which equals the likelihood of the (given) future observations 
conditioned on $\vct{X}(t)$ (up to a scale factor). 

From \Fig{fig:FactorGraph}, we determine these quantities 
using Tables II--IV of \cite{LDHKLK:fgsp2007} as follows. 
From (III.1) and (II.7) of \cite{LDHKLK:fgsp2007}, we have 
\begin{IEEEeqnarray}{rCl} \label{eqn:VfHalfstep1}
\vmsgf{V}{\vct{X}(t_k^-)} &=& 
e^{\vct{A}\Delta} \vmsgf{V}{\vct{X}(t_{k-1})} (e^{\vct{A}\Delta})^\T
 +  \sigma_U^2\Delta \vct{B} \vct{B}^\T,
\end{IEEEeqnarray}
and from (IV.2) and (IV.3) of \cite{LDHKLK:fgsp2007}, we have
\begin{IEEEeqnarray}{rCl} \label{eqn:VfHalfstep2}
\vmsgf{V}{\vct{X}(t_k)} &=& \vmsgf{V}{\vct{X}(t_k^-)} \nonumber \\
&&- \vmsgf{V}{\vct{X}(t_k^-)} \vct{C} \left(\frac{\sigma_Z^2}{\Delta} \vct{I}_{o} + \vct{C}^\T \vmsgf{V}{\vct{X}(t_k^-)} \vct{C}\right)^{-1} \vct{C}^\T \vmsgf{V}{\vct{X}(t_k^-)} \nonumber \\
\end{IEEEeqnarray}
For $\Delta\approx 0$, we have 
\begin{IEEEeqnarray}{rCl} \label{eqn:LinApproxexpADelta}
e^{\vct{A}\Delta} &\approx& \vct{I}_{n} + \Delta \vct{A};
\end{IEEEeqnarray}
thus (\ref{eqn:VfHalfstep1}) becomes 
\begin{IEEEeqnarray}{rCl}
\IEEEeqnarraymulticol{3}{l}{
\vmsgf{V}{\vct{X}(t_k^-)} \approx \vmsgf{V}{\vct{X}(t_{k-1})}
}\nonumber\\\quad
 && {} + \Delta\left( \vct{A}\vmsgf{V}{\vct{X}(t_{k-1})} + (\vct{A}\vmsgf{V}{\vct{X}(t_{k-1})})^\T 
               + \sigma_U^2 \vct{B} \vct{B}^\T \right)
 \IEEEeqnarraynumspace \label{eqn:VfHalfstep1Lim}
\end{IEEEeqnarray}
and (\ref{eqn:VfHalfstep2}) becomes
\begin{IEEEeqnarray}{rCl} \label{eqn:VfHalfstep2Lim}
\vmsgf{V}{\vct{X}(t_k)} &\approx& \vmsgf{V}{\vct{X}(t_k^-)}  
- \frac{\Delta}{\sigma_Z^2} \vmsgf{V}{\vct{X}(t_k^-)} \vct{C} \vct{C}^\T \vmsgf{V}{\vct{X}(t_k^-)}.
\end{IEEEeqnarray}
Combining (\ref{eqn:VfHalfstep1Lim}) and (\ref{eqn:VfHalfstep2Lim}) 
yields (\ref{eqn:DefVf})--(\ref{eqn:ForwRiccati})
as the steady-state condition for 
\begin{IEEEeqnarray}{rCl} \label{eqn:VfromVX}
\vmsgf{V}{} &\eqdef& \vmsgf{V}{\vct{X}(t)}/\sigma_U^2
\end{IEEEeqnarray}
in the limit $\Delta \rightarrow 0$.

The derivation of (\ref{eqn:BackwRiccati}) is essentially identical except that 
the matrix $e^{\vct{A}\Delta}$ is replaced by its inverse, which amounts 
to a sign change in $\vct{A}$.

As for $\vmsgf{m}{\vct{X}(t)}$, we have 
\begin{IEEEeqnarray}{rCl} \label{eqn:mfHalfstep1}
\vmsgf{m}{\vct{X}(t_k^-)} &=& e^{\vct{A}\Delta} \vmsgf{m}{\vct{X}(t_k)} + \vct{\Gamma}\vct{s}(t_{k-1})\Delta
\end{IEEEeqnarray}
from (III.2) and  (II.9) of \cite{LDHKLK:fgsp2007}, and 
\begin{IEEEeqnarray}{rCl}
\vmsgf{m}{\vct{X}(t_k)} & = & \vmsgf{m}{\vct{X}(t_k^-)} \nonumber \\
  &&- \vmsgf{V}{X(t_k^-)} \vct{C} \left(\frac{\sigma_Z^2}{\Delta}\vct{I}_{o} + \vct{C}^\T \vmsgf{V}{X(t_k^-)} \vct{C}\right)^{-1} \vct{C}^\T \vmsgf{m}{\vct{X}(t_k^-)} \nonumber \\%
\end{IEEEeqnarray}
from (IV.1) and (IV.3) of \cite{LDHKLK:fgsp2007}.
For $\Delta\approx 0$, we obtain with (\ref{eqn:LinApproxexpADelta})
\begin{IEEEeqnarray}{rCl}
\IEEEeqnarraymulticol{3}{l}{
\vmsgf{m}{\vct{X}(t_k)} = \vmsgf{m}{\vct{X}(t_{k-1})} 
     + \Delta\Big( \vct{A}\vmsgf{m}{\vct{X}(t_{k-1})}
}\nonumber\\\quad
 {} + \vct{\Gamma}\vct{s}(t_{k-1}) 
  - \frac{1}{\eta^2} \vmsgf{V}{} \vct{C}\vct{C}^\T \vmsgf{m}{\vct{X}(t_{k-1})}  \Big),
 \IEEEeqnarraynumspace\label{eqn:mstepDifferential}
\end{IEEEeqnarray}
where we have used the normalized stationary covariance matrix (\ref{eqn:VfromVX}).
Note that (\ref{eqn:mstepDifferential}) is exact in the limit $\Delta\rightarrow 0$ 
and amounts to the differential equation
\begin{IEEEeqnarray}{rCl}
\frac{d}{dt} \vmsgf{m}{\vct{X}(t)} 
  &=& \left(\vct{A} - \frac{1}{\eta^2} \vmsgf{V}{} \vct{C}\vct{C}^\T \right) \vmsgf{m}{\vct{X}(t)} 
    + \vct{\Gamma}\vct{s}(t).
\end{IEEEeqnarray}
The solution of this differential equation (for $t>0$) is 
\begin{IEEEeqnarray}{rCl}
\vmsgf{m}{\vct{X}(t)} &=& e^{\vct{\tilde{A}}t} \vmsgf{m}{\vct{X}(0)} 
 + e^{\vct{\tilde{A}}t} \int_{0}^t e^{-\vct{\tilde{A}}\tau} \vct{\Gamma}\vct{s}(\tau)\, d\tau
\end{IEEEeqnarray}
with
$\vct{\tilde{A}} \eqdef \vct{A} - \vmsgf{V}{} \vct{C}\vct{C}^\T / \eta^2$. 
This solution applies to any interval between $t_k$ and $t_{k+1}$ 
in Section~\ref{sec:Recursions} and yields (\ref{eqn:ForwardRecursion}) 
with (\ref{eqn:Af}) and~(\ref{eqn:Bf}).

The derivation for $\vmsgb{m}{\vct{X}(t)}$ is essentially identical except
for a sign change in both $\vct{A}$ and $\vct{\Gamma}$, 
where the latter is due to (II.10) of \cite{LDHKLK:fgsp2007}.

Finally, we use the result from \cite{BLV:ctfg2013arXiv} that 
the MAP/MMSE/LMMSE estimate of $U(t)$ 
(i.e., the posterior mean of (\ref{eqn:DefLocalAverageUt}) for $\Delta\rightarrow 0$)
is given by
\begin{IEEEeqnarray}{rCl} \label{eqn:InputEstimate}
\hat{\vct{u}}(t) &=& 
\sigma_U^2 \vct{B}^\T 
\tilde{\vct{W}}(t)
\left( \vmsgb{m}{X(t)} - \vmsgf{m}{X(t)} \right)
\end{IEEEeqnarray}
with 
\begin{IEEEeqnarray}{rCl} \label{eqn:tildeW}
\tilde{\vct{W}}(t) &\eqdef& \left( \vmsgf{V}{\vct{X}(t)} + \vmsgb{V}{\vct{X}(t)} \right)^{\!-1},
\end{IEEEeqnarray}
which yields (\ref{eqn:uhatcomp}) and~(\ref{eqn:wEquation}). 
Note that (\ref{eqn:InputEstimate}) and (\ref{eqn:tildeW}) may also be obtained 
directly from \Fig{fig:FactorGraph} using 
(II.12), (III.8), and (III.9) of \cite{LDHKLK:fgsp2007} 
and then taking the limit $\Delta\rightarrow 0$.

\section*{Acknowledgment}

The authors would like to thank Jonas Biveroni 
and Patrik Strebel for building the hardware prototype 
of Section~\ref{sec:HardwareMeasurements}.
The helpful comments by Hanspeter Schmid 
are also gratefully acknowledged.

\newcommand{\CASI}{IEEE Trans.\ Circuits \& Systems~I}
\newcommand{\CASII}{IEEE Trans.\ Circuits \& Systems~II: Express Briefs}
\newcommand{\COM}{IEEE Trans.\ Communications}
\newcommand{\COMMag}{IEEE Communications Mag.}
\newcommand{\IT}{IEEE Trans.\ Information Theory}
\newcommand{\JSAC}{IEEE J.\ Select.\ Areas in Communications}
\newcommand{\SP}{IEEE Trans.\ Signal Proc.}
\newcommand{\SPMag}{IEEE Signal Proc.\ Mag.}
\newcommand{\ProcIEEE}{Proceedings of the IEEE}
\newcommand{\ESTCS}{IEEE J.\ Emerg.\ and\ Select.\ Topics in Circuits \& Systems}

\end{document}